\documentclass[remotesensing,article,accept,moreauthors]{Definitions/mdpi}

\usepackage{xspace}

\newcommand{\Rtwo}{\ensuremath{R^2}\xspace}

\graphicspath{{figures/}}
\newcommand{\AllProductsGrassProductRows}{13,506\xspace}
\newcommand{\AllProductsGrassEvents}{5741\xspace}
\newcommand{\AllProductsGrassRtwo}{0.715\xspace}
\newcommand{\AllProductsGrassRmse}{44.7\xspace}

\newcommand{\AllProductsShrubProductRows}{67,437\xspace}
\newcommand{\AllProductsShrubEvents}{25,340\xspace}
\newcommand{\AllProductsShrubRtwo}{0.693\xspace}
\newcommand{\AllProductsShrubRmse}{23.3\xspace}

\newcommand{\AllProductsTreeProductRows}{37,976\xspace}
\newcommand{\AllProductsTreeEvents}{15,062\xspace}
\newcommand{\AllProductsTreeRtwo}{0.691\xspace}
\newcommand{\AllProductsTreeRmse}{25.6\xspace}

\newcommand{\NoModisGrassRtwo}{0.645\xspace}

\newcommand{\NoModisShrubRtwo}{0.680\xspace}

\newcommand{\NoModisTreeRtwo}{0.689\xspace}

\firstpage{1}
\pubvolume{18}
\issuenum{21}
\articlenumber{3744}
\pubyear{2026}
\copyrightyear{2026}
\externaleditor{Carmen Quintano}
\datereceived{3 September 2026}
\daterevised{15 October 2026}
\dateaccepted{26 October 2026}
\datepublished{30 October 2026}

\Title{A Unified Multisensor Machine-Learning Framework for Live Fuel Moisture Content Retrieval}

\Author{ {Valerio Pampanoni} %MDPI: Please carefully check the accuracy of names and affiliations. @VP: I confirm that all the author names, affiliations and ORCIDs are correct.
 $^{1,}$*\orcidA{}, Emilio Chuvieco $^{2}$\orcidB{},
Alvise Ferrari $^{3}$\orcidC{}, Giovanni Laneve $^{1}$\orcidD{} and
Simone Saquella $^{4}$\orcidE{}}

\AuthorNames{Valerio Pampanoni, Emilio Chuvieco, Alvise Ferrari, Giovanni Laneve and Simone Saquella}

\address{%
$^{1}$ \quad School of Aerospace Engineering, Sapienza University of Rome, 00138 Rome, Italy %MDPI: We added these email addresses here according to those submitted online at susy.mdpi.com. Please confirm. @VP: actually, only my email should appear on the paper, all correspondence should be addressed to valerio.pampanoni@uniroma1.it. The other email addresses have now been removed from the affiliations.
\\
$^{2}$ \quad Environmental Remote Sensing Research Group, Department of Geology, Geography and the Environment, Universidad de Alcal\'a, 28801 Alcal\'a de Henares, Spain\\
$^{3}$ \quad GMATICS S.r.l., 00173 Rome, Italy\\
$^{4}$ \quad Serco Italia S.p.A., 00173 Rome, Italy}

\corres{Correspondence: valerio.pampanoni@uniroma1.it}

\addhighlights{yes}
\renewcommand{\addhighlights}{%
\par\noindent
\textbf{What are the main findings?}
\begin{itemize}[labelsep=2.5mm,topsep=-3pt]
\item Harmonised optical observations from six satellite products can share vegetation-specific models for estimating live fuel moisture content.
\item Multisensor training reduces prediction errors in all 18 product--vegetation comparisons when tested on withheld dates at represented sites.
\end{itemize}\vspace{3pt}
\textbf{What are the implications of the main findings?}
\begin{itemize}[labelsep=2.5mm,topsep=-3pt]
\item A common retrieval framework can serve local fuel-moisture mapping and regional monitoring while retaining each satellite product's spatial resolution.
\item The framework can accommodate additional optical sensors when compatible bands and overlapping observations support calibration and validation.
\end{itemize}}

\abstract{Live fuel moisture content (LFMC) influences vegetation flammability, but field measurements are too sparse to describe its variation over large areas. We present, to our knowledge, the first LFMC framework that combines observations from multiple optical satellite sensors within the same trained models. More than two decades of GlobeLFMC 2.0 measurements are matched to Terra and Aqua Moderate Resolution Imaging Spectroradiometer (MODIS), Visible Infrared Imaging Radiometer Suite (VIIRS), Landsat 8/9, Sentinel-2 and Sentinel-3. We screen the vegetation surrounding sampling locations and adjust reflectance for differences between sensors before calculating vegetation indices. Weather, topography and time of year provide additional information. Separate models are trained for grasses, shrubs and trees. We test prediction on dates not used for training at sites where other dates contribute to training. At each site, records using the same satellite image are kept in either the training or the test set. The final Random Forest models achieve $R^2$ values of \AllProductsGrassRtwo, \AllProductsShrubRtwo and \AllProductsTreeRtwo, respectively. Models trained on the combined archive give higher $R^2$ and lower root mean square error than models trained on one satellite product in all \mbox{18 product}--vegetation comparisons on identical test observations. Paired 95\% confidence intervals support the improvement in 16 comparisons. Testing at sites excluded from training gives lower accuracy. The method produces LFMC maps at the selected satellite product's resolution, from 20--30~m to 300--500~m.\vspace{-2pt}}


\begin{document}

\section{Introduction}

Live fuel moisture content (LFMC) is the mass of water in living vegetation
relative to dry biomass, expressed as a percentage. It is one of the most
direct fuel-state controls on ignition probability, flame spread and fire
behaviour \citep{chuvieco2004, yebra2013review, chuvieco2023towards}. Fire
danger rating and early-warning systems are normally applied over regional to
continental domains and updated repeatedly through the fire season; they
therefore require spatially continuous fuel-condition information rather than
isolated field measurements \citep{usgs2026wfpi, laneve2020daily,
chuvieco2023towards}. Field LFMC measurements are accurate but sparse,
temporally irregular and concentrated in a limited number of countries and
ecosystems. While optical satellite products enable spatially frequent and repeatable 
monitoring of canopy reflectance and its derived spectral indices, their relationship 
with the water and dry mass content of live vegetation remains complex. Retrieval therefore requires an
empirical or physical relationship between field-measured LFMC and remotely
sensed, meteorological and environmental predictors. A large-area LFMC product
intended for repeated fire-danger applications must also keep this
relationship consistent across sensors, spatial supports and acquisition
periods.

Historically, remote-sensing LFMC retrieval techniques can be classified into two major families: empirical methods
and radiative-transfer model (RTM) inversion \citep{quan2021}. Empirical models relate field LFMC to spectral vegetation indices, thermal information,
meteorological variables and other ancillary information \citep{yebra2008}. Recently, multivariate machine-learning models combining several predictor
families have emerged \citep{cunill2022rf, yebra2026highres, wang2026tcn}.
Some approaches also incorporate physical constraints \citep{rao2020sar}. Their main advantage is
flexibility: they can combine optical, meteorological, topographic and
seasonal information and can exploit large field archives directly. Their main
limitation is that extrapolation outside the training domain depends on how
well the sampled sites, sensors and environmental conditions represent the
application domain. Physically-based approaches instead invert canopy
radiative-transfer models, often using lookup tables from PROSAIL, which couples the PROSPECT leaf
model with the scattering by arbitrarily inclined leaves (SAIL) canopy
model, constrained by
ecological priors \citep{yebra2009linking, quan2021, pampanoni2022s3}. This strategy has a
clearer mechanistic basis and can expose which leaf and canopy parameters are
consistent with an observed spectrum. Hybrid RTM--machine-learning strategies
have also recently been proposed to preserve a physical simulation basis while
using statistical regressors for efficient LFMC prediction
\citep{pampanoni2024prosail}. However, LFMC inversion is severely ill-posed:
different combinations of leaf water, dry matter, canopy structure,
background and geometry can generate similar reflectance spectra. Ecological
constraints help reduce this ambiguity by restricting the combinations of leaf
and canopy properties considered during inversion \citep{riano2005,
yebra2009linking, quan2021, pampanoni2023thesis, pampanoni2024prosail}. 

Most LFMC retrieval studies remain tied to a single satellite product or
closely related product family. Examples include empirical and
physical-inversion products based on the Moderate Resolution Imaging
Spectroradiometer (MODIS) \citep{yebra2008, quan2021}, regional MODIS
machine-learning approaches \citep{cunill2022rf}, and high-resolution
Sentinel-2 products \citep{yebra2026highres}. Passive-microwave observations
provide a complementary route by relating LFMC to retrieved vegetation water
content and soil moisture at much coarser spatial support \citep{cho2025smap}.
Single-product designs simplify
calibration because the spectral response functions (SRFs), spatial support,
compositing rules and quality masks are fixed. They also have the drawback
of limiting the amount of GlobeLFMC observations usable for validation to the
timeline of the target sensor, which in turn severely limits the variety of
climatic zones, biomes, vegetation species and observations available for training
and validation. Multisensor modelling can address this limitation, but only if product differences are treated
explicitly. A model trained on pooled observations can
otherwise confound LFMC-related variability with product artefacts introduced
by sensor architecture, footprint size or compositing.

This study focuses on empirical machine-learning retrieval and addresses two
linked needs. First, we aim to make fuller use of GlobeLFMC 2.0, the recently
updated version of the largest field LFMC archive \citep{yebra2024globelfmc}.
Combining satellite products with different observation periods increases the
number of field measurements that can be matched to usable imagery and broadens
the sampled biomes, environmental conditions and species available for training
and validation, subject to product-specific suitability and quality criteria.
The long MODIS record is particularly valuable because it includes field dates
preceding the newer missions and enables historical LFMC analyses over more
than two decades. The benefit of this extended coverage is considered
separately from the predictive contribution of MODIS training observations.

Second, we aim to provide the remote-sensing and fire-management communities
with a common LFMC estimation method that can be used at different spatial
resolutions. Retaining each product's native spatial support allows the
satellite input to be selected according to the spatial detail and temporal
coverage required by an application. To our knowledge, this is the first LFMC
framework designed to combine these two goals within a common trained
multisensor method. The implications for local fuel monitoring and broader
fire-danger applications are discussed alongside the results.

To this end, we propose a unified framework combining satellite, meteorological,
topographic and seasonal predictors. We match GlobeLFMC 2.0 field records to
Terra MODIS MOD09A1, Aqua MODIS MYD09A1, Visible Infrared Imaging Radiometer
Suite (VIIRS) VNP09H1, Landsat 8/9 Collection 2 Level-2, Sentinel-2 Multispectral
Instrument (MSI) Level-2A and Sentinel-3 SY\_2\_SYN 300 m observations. The
common optical predictors use red, near-infrared (NIR) and shortwave-infrared
reflectance at approximately 1.6~$\upmu$m (SWIR1). Product-specific site suitability and
reflectance calibration account for differences in spatial support and spectral
response before vegetation indices are combined with environmental predictors;
their implementation is described in Section~\ref{sec:methods}. We fit separate
models for grasses, shrubs and trees (hereafter Grass, Shrub and Tree), which
differ in canopy structure, rooting strategies, phenological patterns and LFMC
dynamics~\mbox{\citep{yebra2013review, quan2021, yebra2024globelfmc}.}

We evaluate LFMC estimation for additional dates at sites represented in
training and determine whether pooling products improves prediction within
each product's observations compared with product-specific training on identical
held-out records. A controlled comparison additionally evaluates the contribution
of Terra and Aqua MODIS training observations on the same non-MODIS test records.
The common predictor definitions provide a basis for incorporating future sensors when compatible reflectance bands, documented spectral responses and sufficient overlapping observations are available for calibration and~\mbox{validation}.

The paper is organised as follows. Section~\ref{sec:data} describes
GlobeLFMC 2.0, the satellite products and the ancillary variables used as
predictors. Section~\ref{sec:methods} explains the preliminary feature-set
screening, unified sampling, site-suitability filtering, SRF diagnostics,
empirical reflectance calibration, machine-learning models and validation
design.
Section~\ref{sec:results} reports the feature-screening outcome, filtered
sample support, harmonisation diagnostics, within-site LFMC retrieval skill,
and predictor-importance diagnostics.
Section~\ref{sec:discussion} discusses the role of MODIS, the conditions under
which the sensor-agnostic interpretation is valid, and the implications for
LFMC mapping at different spatial scales.
Appendix~\ref{app:site-blocked-validation} presents complementary site-blocked
validation; Appendix~\ref{app:camp-fire-case} illustrates pre-fire LFMC mapping
for the Camp Fire, including seasonal maps and a Sentinel-2/Sentinel-3 comparison;
\mbox{Appendix~\ref{app:multisensor-diagnostics}} reports the controlled MODIS ablation
and matched product-specific versus multisensor benchmarks, including
product-wise optical increments under both training regimes; and
Appendix~\ref{app:filter-sensitivity} examines sensitivity to the site-suitability
and LFMC-variability filters. \mbox{Appendix~\ref{app:sentinel1-optical-free}} evaluates
an optical-free Sentinel-1 branch against an environmental-only~\mbox{baseline}.

\section{Materials and Methods}
\label{sec:materials-methods}

\begin{figure}[H]
\includegraphics[width=0.99\textwidth]{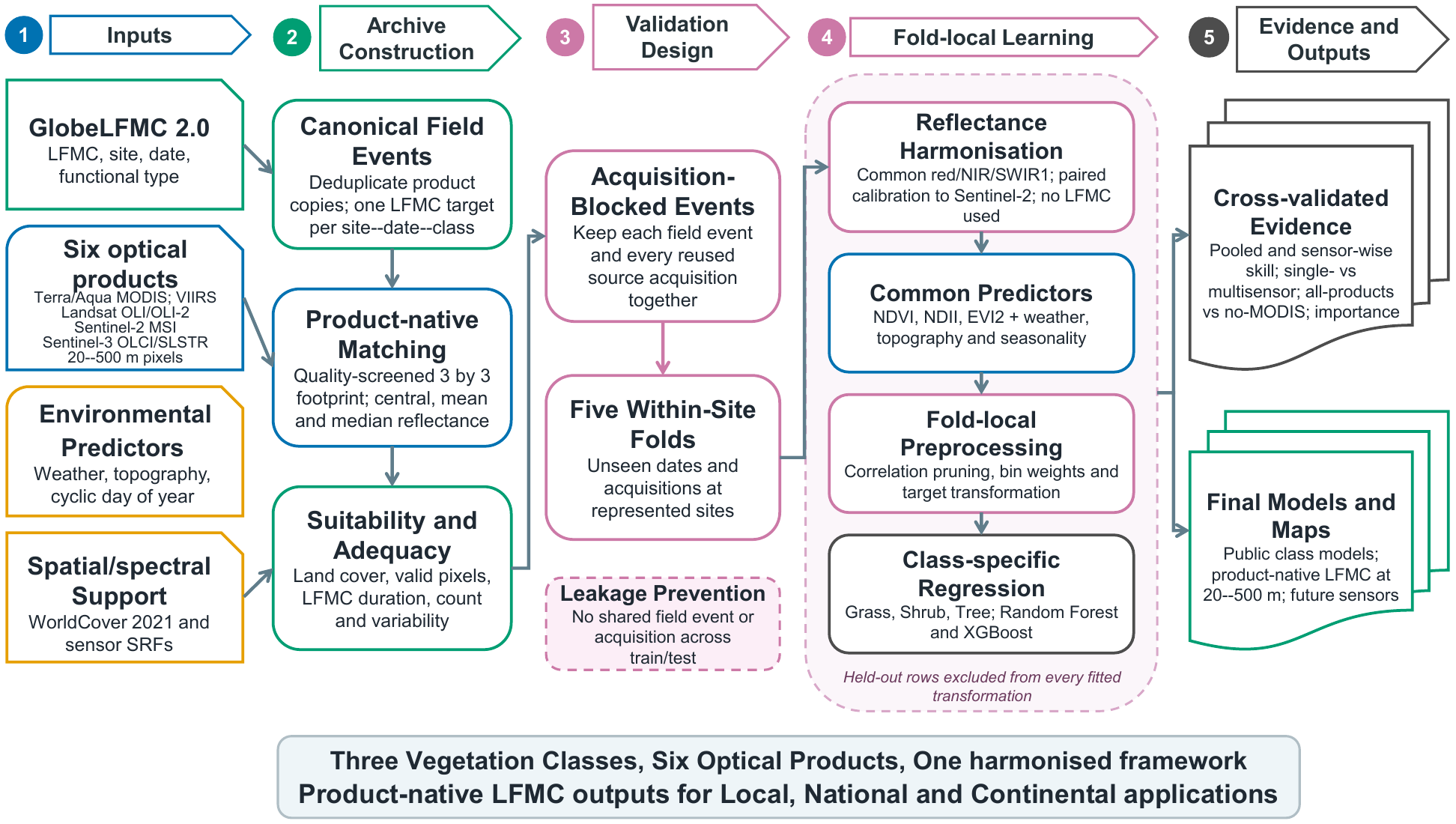}
\caption{Overview of the unified multisensor LFMC workflow. GlobeLFMC 2.0
measurements are converted to canonical site--date--functional-class events and
matched to quality-screened, product-native satellite footprints.
Product-specific land-cover suitability and LFMC time-series adequacy criteria
are applied before validation. Field events and reused source acquisitions are
grouped into indivisible validation blocks. Within each outer fold,
target-independent reflectance calibration, predictor preprocessing and model
fitting exclude the held-out rows. The resulting Grass, Shrub and Tree models
share a common predictor space and can generate LFMC estimates at the native
spatial support of the selected satellite product.}
\label{fig:workflow}
\end{figure}

Figure~\ref{fig:workflow} summarises the complete workflow from canonical
field-event construction and product-native satellite sampling to
leakage-controlled multisensor training, evaluation and final LFMC mapping.

\subsection{Data Sources}
\label{sec:data}

\subsubsection{GlobeLFMC 2.0  {In Situ} %MDPI: We removed the italic format of the Latin Expression in the full text. Please confirm. @VP: you're right, latin text should not be italicized, thanks for correcting this.
 Database}

\begin{figure}[H]
\includegraphics[width=0.99\textwidth]{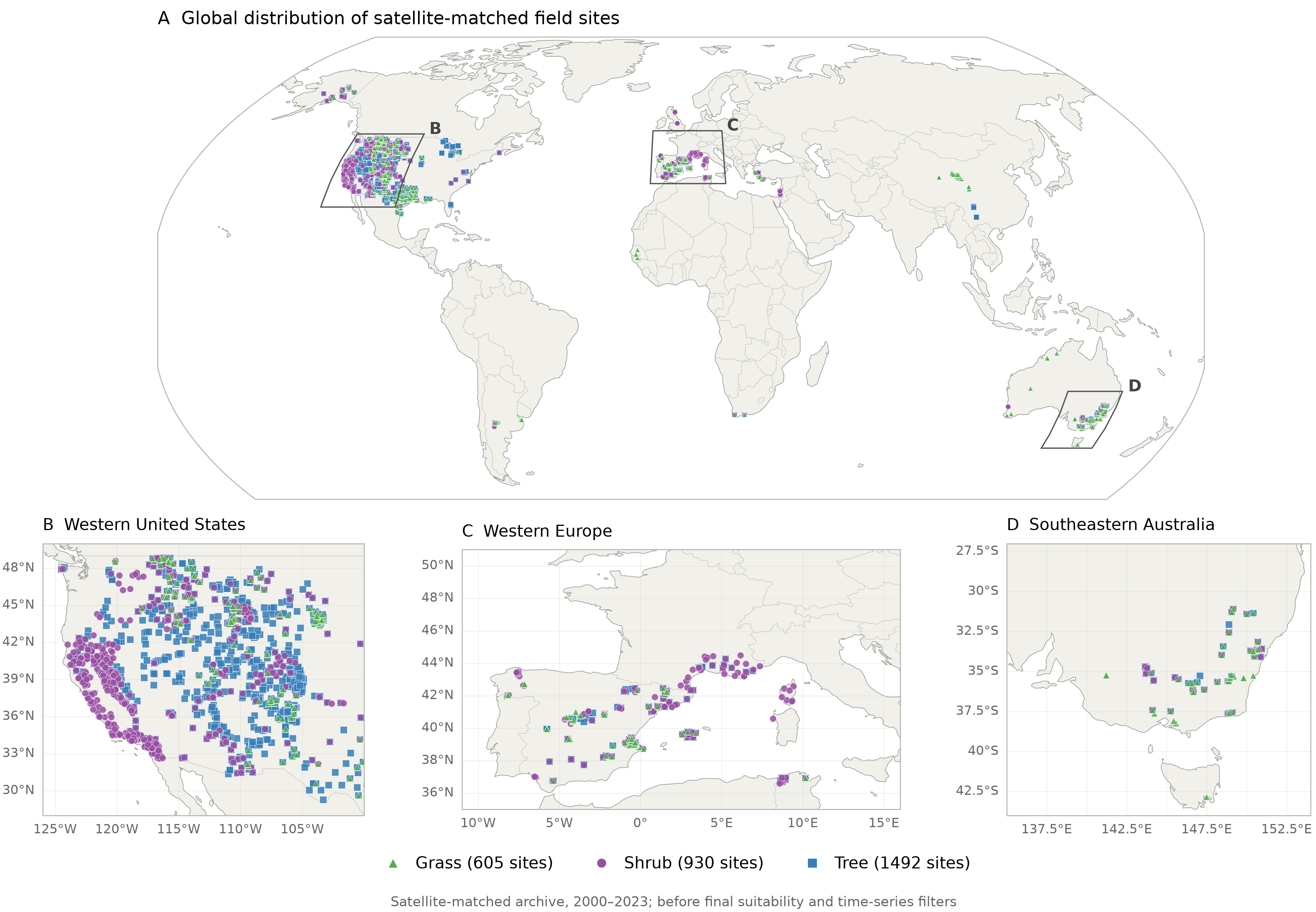}
\caption{ {GlobeLFMC} %MDPI: 1. Please change the hyphen (-) between numbers into an en dash (−, “U+2013”), e.g., “2019-2022” should be “2019–2022” and “0-9” should be “0–9”. 2. Commas are only used for numbers with five or more digits. Please remove them from the four-digit numbers in the figure, e.g., “1,200” should be “1200”. @VP: Figure 2 checked and revised: numerical ranges use en dashes and the four-digit site count has no thousands comma.
 2.0 sites represented in the unified multisensor
satellite-matched archive. Panel (\textbf{A}) shows the global distribution; panels (\textbf{B}--\textbf{D}) provide enlarged views of the western United States, western Europe and southeastern Australia, respectively. Colours and marker shapes distinguish Grass, Shrub and Tree sites. Points show unique site--functional-class combinations retained after matching field records to the availability windows of Terra MODIS MOD09A1, Aqua MODIS MYD09A1, VIIRS VNP09H1, Landsat 8/9 Collection 2 Level-2, Sentinel-2 MSI Level-2A and Sentinel-3 SY\_2\_SYN 300 m. The map is intended to show the geographic
distribution of the field archive available to the multisensor workflow before
the final site-suitability, time-series adequacy and valid-\mbox{pixel filters}.}
\label{fig:unified-sites}
\end{figure}

\begin{figure}[H]
\includegraphics[width=\textwidth]{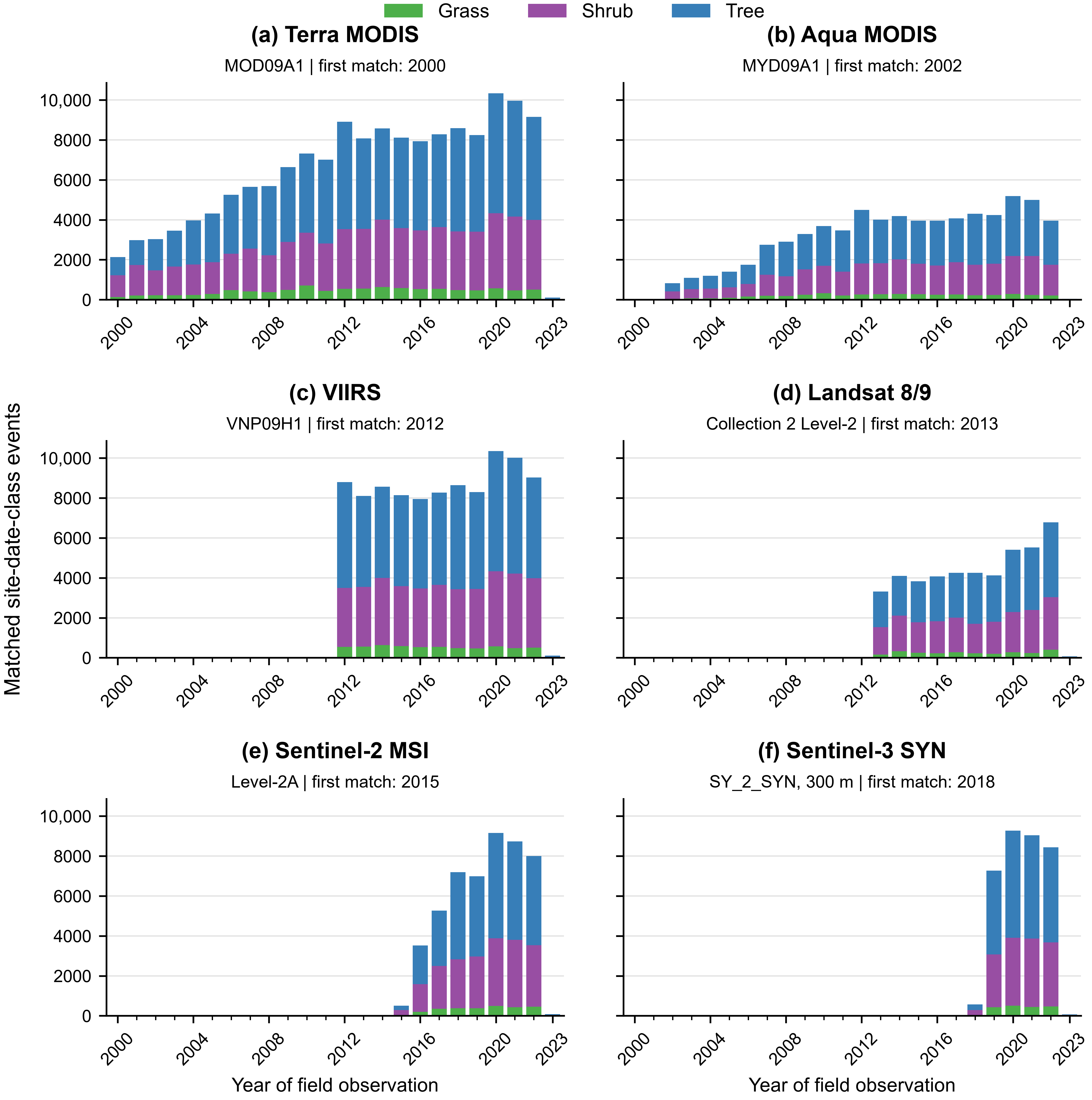}
\caption{ {Satellite-matched} %MDPI: Commas are only used for numbers with five or more digits. Please remove them from the four-digit numbers in the figure, e.g., “1,200” should be “1200”. @VP: Figure 3 revised: four-digit tick values have no thousands commas; five-digit values retain commas.
 GlobeLFMC 2.0 observations available to the unified
multisensor sampling table from 2000 to 2023. Bars show unique
site-date-functional-class events with a matched acquisition and valid central-pixel
red, near-infrared and shortwave-infrared reflectances, before the final
site-suitability and time-series filters, separated by satellite product and
vegetation class. Years refer to field-observation dates; unmatched sampling
requests are excluded. The
Terra MODIS MOD09A1 branch is the only one covering the full satellite-matched
period, while Aqua MODIS MYD09A1, VIIRS VNP09H1, Landsat 8/9 Collection 2
Level-2, Sentinel-2 MSI Level-2A and Sentinel-3 SY\_2\_SYN add observations after their availability
dates.}
\label{fig:unified-year-counts}
\end{figure}

GlobeLFMC 2.0 is the  {in situ} reference dataset used to train and evaluate the
models~\citep{yebra2024globelfmc}. It contains field LFMC measurements, site
coordinates, dates, taxonomic metadata, functional vegetation classes,
quality flags, land-cover descriptors and meteorological variables. The
complete archive spans 1977--2023, but the satellite-matched
archive used in this study begins in 2000 because Terra MODIS MOD09A1 is the
earliest supported reflectance product. Figure~\ref{fig:unified-sites} maps
the GlobeLFMC sites represented in the unified satellite-matched archive, and
Figure~\ref{fig:unified-year-counts} shows the observations available to each
product after matching the field records to the product availability windows.

The GlobeLFMC 2.0 fields used in the workflow are summarised in
Table~\ref{tab:globelfmc-fields}. The site name defines grouping for
site-level filtering and validation. Latitude and longitude are used for
spatial matching but are not model predictors. Sampling date is used to match
satellite and weather observations and to derive cyclic day-of-year variables.
Sampling time is not used because it is unavailable for most of the records.
GlobeLFMC combines destructive plant-material measurements from multiple
monitoring programmes, with sampling and weighing procedures documented in
its Protocol sheet \citep{yebra2024globelfmc}. It has no uniform field-plot
resolution: site coordinates locate the sampling sites, but the standard
metadata do not specify plot dimensions. The product-native satellite
footprints described below therefore represent the remote-sensing support,
not the dimensions of the field plots.
Species functional type is harmonised into three modelling classes: Grass, Shrub and
Tree. Parenthetical counts below refer to individual field records in the complete
GlobeLFMC 2.0 source database, including records collected before 2000. No date
restriction, satellite matching, field-quality, site-suitability or time-series
adequacy filter is applied to these counts; each field record is counted once.
The Grass class includes records labelled as Grass (12,635), Forb (1349),
Graminoid (91), Sedge (185) or Geophyte (4); the Shrub class includes Shrub
(110,848), Large shrub (419), Subshrub (171) and Shrub/Small tree (526); and the
Tree class includes Tree (166,936), Small tree (443) and Liana (25).
GlobeLFMC does not provide mapped field-plot polygons or fractional cover
of Grass, Shrub and Tree within each satellite footprint; sites may therefore
include measurements from one or more sampled species or functional types, but
not a satellite-ready map of their areal cover. Species names and other metadata
are retained for auditability but are not used as predictors in the model.
These modelling classes describe broad functional groupings rather than
individual biomes. In particular, Tree does not distinguish boreal,
temperate or other forest environments, species composition or canopy
structure; these sources of variation remain within the pooled class.

Some GlobeLFMC site--date records contain multiple field measurements. These may
be replicate samples of the same species, measurements of several species within
the same functional class, or measurements of different functional classes at
the same site and date. Field-level quality rules, including removal of records
identified as isolated measurements, are applied before constructing the
event-level LFMC targets used for model training and evaluation. A
canonical field-measurement identifier is then formed independently of the
satellite products, and repeated copies introduced when one field measurement is
matched to several products are removed. The remaining measurements are
aggregated once within each site, date and functional class using their arithmetic
mean. This event-level mean LFMC target is used consistently across vegetation
classes, product scenarios and model comparisons, giving every retained field
measurement equal contribution within its event. No single species measurement
is selected arbitrarily when multiple same-class measurements exist. The
outcome definition was standardised during manuscript revision; it is fixed in
the revised training grid and is not selected by predictive performance.
Exploratory mean-versus-median target comparisons are retained in Supplementary
File S5. The two aggregations define different outcomes, and differences between
their scores are reported as descriptive sensitivity results.

\vspace{-6pt}

\begin{table}[H]
\caption{GlobeLFMC 2.0 fields used in the unified multisensor LFMC pipeline. Latitude and longitude are used only for spatial matching and are not model predictors.}
\label{tab:globelfmc-fields}

\begin{adjustwidth}{-\extralength}{0cm}
%\centering %% If there is a figure in wide page, please release command \centering, for Table, ``\textwidth" should be ``\fulllength"
\begin{tabularx}{\fulllength}{@{\extracolsep{\fill}}>{\raggedright\arraybackslash}m{0.18\fulllength}>{\raggedright\arraybackslash}m{0.27\fulllength}>{\raggedright\arraybackslash}m{0.48\fulllength}@{}}
\toprule
\textbf{Field Group} & \textbf{Variables} & \textbf{Use in This Study} \\
\midrule
Site identity and location & Site name; country; state/region; latitude; longitude & Defines sampling locations, site grouping, geographic diagnostics and validation groups; coordinates are not predictors. \\\midrule
Sampling time & Sampling date; sampling time where available & Sampling date is used to match satellite and weather observations and to derive cyclic day-of-year predictors; time of day is not used because it is mostly unavailable. \\\midrule
Vegetation information & Species collected; species functional type; old/new leaves & Functional type is harmonised to Grass, Shrub and Tree; species metadata supports aggregation audits but is not used as a predictor. \\\midrule
LFMC target & LFMC value (\%) & After field-level quality screening, canonical field measurements are deduplicated across products and aggregated once to one mean or median target per site-date-functional-class event; the same target is then assigned to every matched product row. \\\midrule
Quality/context flags & Isolated data point; individual/mean flag; extra information; quality flag & Isolated observations are removed; remaining flags are retained for audit and interpretation. \\\midrule
Native ancillary data & AgERA5 weather; International Geosphere-Biosphere Programme (IGBP) land cover; elevation; slope & Weather variables are used as predictors; IGBP is diagnostic only; topography is resampled consistently from the Advanced Land Observing Satellite World 3D \mbox{30 m} digital surface model. \\
\bottomrule
\end{tabularx}
\end{adjustwidth}
\end{table}

A field event can be matched to more than one satellite product. These matches
are retained as separate product-observation rows because each contains a
distinct spectral measurement needed to train a model that operates across
sensors. They are not treated as independent LFMC measurements: the canonical
event-level target is calculated before the product tables are joined and the
same value is attached to every product row associated with that
site--date--functional-class event. Assertions in the data-construction and
training code reject any event for which product-dependent target values remain.
Their treatment during fitting and validation is described in
Sections~\ref{sec:ml-models} and \ref{sec:validation-design}.

\begin{figure}[H]
\includegraphics[width=0.99\textwidth]{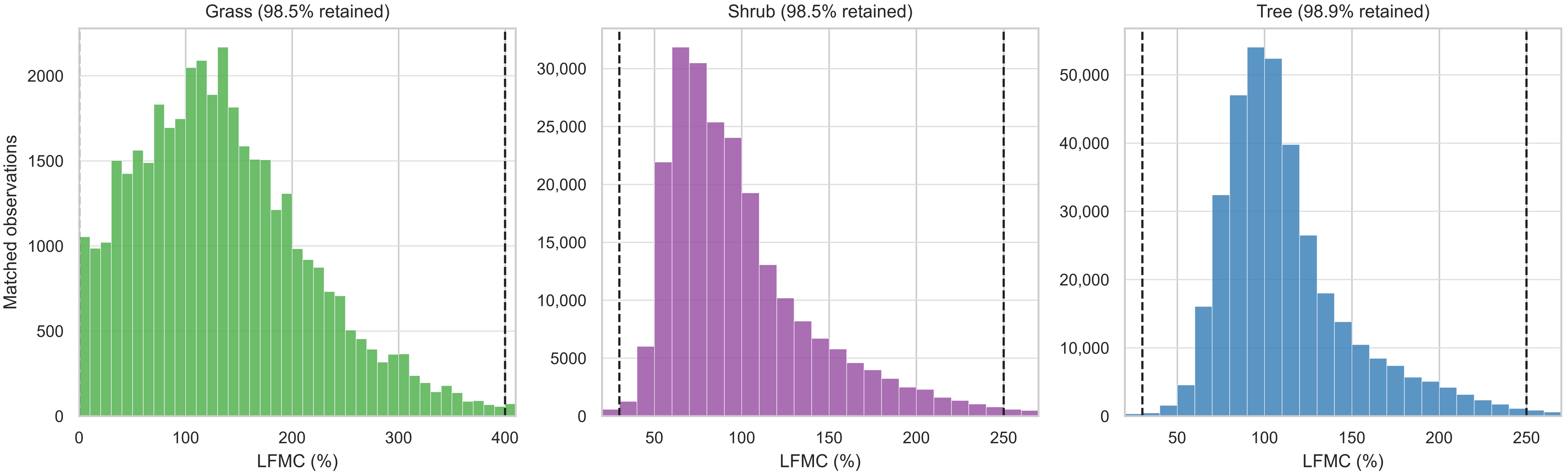}
\caption{ {LFMC} %MDPI: Please use commas to separate out the thousands in numbers with five or more digits (but not four digits) in the picture, e.g., “10000” should be “10,000”. @VP: Figure 4 revised: values with five or more digits use thousands commas; four-digit values do not.
 distributions in the unified satellite-matched archive by
functional vegetation class. Dashed vertical lines mark the target modelling
ranges: 0--400\% for Grass and 30--250\% for Shrub and Tree. The retained
fractions printed in each panel are computed from the matched multisensor
archive used by the unified branch.}
\label{fig:unified-lfmc-dist}
\end{figure}

Field LFMC values are restricted to the target LFMC domain supported by the
data distribution: Grass is retained over the 0--400\% LFMC range, while Shrub and Tree are
retained over the 30--250\% range. The broader Grass range is needed because herbaceous
fuels have a much wider high-moisture tail \citep{yebra2013review,
yebra2024globelfmc}. The woody classes have fewer
well-supported observations above 250\%, and those very wet values are less
central to the fire-danger use case. Figure~\ref{fig:unified-lfmc-dist} shows
the matched LFMC distributions and the fraction retained by these bounds.

The lower number of Aqua MODIS matches in Figure~\ref{fig:unified-year-counts}
also reflects reduced availability of valid reflectances. Aqua MODIS has
non-functional detectors in band~6, the approximately 1.6~$\upmu$m SWIR1 band
required by our common spectral indices; affected pixels contain fill values
\citep{nasa_mcst_aqua_detectors}. Over the shared period beginning on
4 July 2002, usable central-pixel matches were available for 97.7\% of the
150,939 candidate field events for Terra and 46.2\% for Aqua. Missing SWIR1
reflectances therefore contribute to Aqua's lower coverage, in addition to
its later start.

\subsubsection{Satellite Reflectance Products}

The unified branch uses six satellite reflectance products
(Table~\ref{tab:satellite-products}). Terra MODIS MOD09A1 Collection 6.1
provides 8-day 500 m surface reflectance from February 2000, and
Aqua MODIS MYD09A1 Collection 6.1 provides the corresponding Aqua MODIS record
from July 2002 \citep{huete2002modis, nasa_mod09a1_061, nasa_myd09a1_061}.
VIIRS VNP09H1 Collection 2 provides 8-day 500 m VIIRS surface reflectance from
January 2012 \citep{vermote2019vnp09, nasa_vnp09h1_002}. Landsat 8 and Landsat 9
Collection 2, Level-2, Tier 1 datasets provide 30 m surface reflectance; the
sampling cutoffs are April 2013 and October 2021, respectively, while Sentinel-2 MSI Level-2A provides
surface-reflectance observations from 2015 in the reprocessed archive~\citep{drusch2012sentinel2}.~All of these datasets are available in Google Earth Engine
and were sampled using the  {official Earth Engine Python client library (version 1.7.35)}%MDPI: Please state which version of the software was used. @VP: Added Earth Engine Python API version 1.7.35, verified against our current environment and its conda installation history.
~\citep{gorelick2017gee}.
Finally, Sentinel-3 SY\_2\_SYN 300 m observations, with field-date sampling
eligibility beginning on 1 May 2018, were located through the Copernicus Data Space Ecosystem (CDSE) OData catalogue and read from its S3 object storage, and add an
additional moderate-resolution Synergy branch based on Ocean and Land Colour Instrument (OLCI) and Sea and Land Surface
Temperature Radiometer (SLSTR) reflectance channels~\citep{copernicus2023s3syn,cdse2026odata}. 
Landsat and Sentinel-2 observations are sampled directly from their respective
Collection 2 Level-2 and MSI Level-2A products. Although the Harmonized Landsat
and Sentinel-2 surface-reflectance products provide a pre-harmonised alternative~\citep{claverie2018hls}, direct sampling preserves the native 30 m Landsat and
20 m Sentinel-2 implementations and allows mission-specific quality screening
and reflectance calibration to be performed within the \mbox{present workflow}.

Field--satellite temporal matching uses product-specific search windows.
For a field date $d$, Sentinel-2 and Landsat candidates are searched from
$d-5$ days at 00:00 Coordinated Universal Time (UTC) up to, but excluding, $d+5$ days at 00:00 UTC.
MODIS and VIIRS candidates are searched over the calendar dates $d-8$ to
$d+8$, inclusive, using the 8-day composite timestamp. The closest eligible
observation is selected after pixel-quality screening and an initial
requirement of at least three valid pixels and 50\% valid footprint support.
For Sentinel-3, scenes within five calendar days of the field date are
ranked by absolute date difference and product name; at most the two nearest
candidates are examined, and the closest meeting the same initial support
requirements is retained. The stricter observation-level thresholds applied
before modelling are described below. Missing or ineligible matches are
excluded. MODIS/VIIRS matching does not use the per-pixel source acquisition
day or require the field date to fall inside the selected composite period;
the recorded offset describes the composite timestamp. Supplementary File S3
specifies the date endpoints, reference-time convention, selection rules and
product-specific quality screening needed to reproduce the sampling table.

\begin{table}[H]
\caption{Satellite reflectance products used by the unified multisensor models. Band identifiers are listed in red/NIR/SWIR1 order. The Landsat 8/9 Operational Land Imager instruments (OLI/OLI-2) use B4/B5/B6, supplied as the Collection 2 surface-reflectance fields SR\_B4/SR\_B5/SR\_B6. Sentinel-3 uses OLCI Oa08/Oa17 and the SLSTR nadir-view band S5N. Footprint width is the square sampling window around each GlobeLFMC site and corresponds to 3-by-3 native-grid pixels.}
\label{tab:satellite-products}

\begin{adjustwidth}{-\extralength}{0cm}
%\centering %% If there is a figure in wide page, please release command \centering, for Table, ``\textwidth" should be ``\fulllength"
\begin{tabularx}{\fulllength}{>{\raggedright\arraybackslash}m{4.5cm}>{\raggedright\arraybackslash}m{4cm}>{\raggedright\arraybackslash}m{2.5cm}>{\raggedright\arraybackslash}m{2.5cm}>{\raggedright\arraybackslash}m{3cm}}
\toprule
\textbf{Product} & \textbf{Record} & \textbf{Native Support} & \textbf{Sampled Footprint} & \textbf{Band IDs} \\
\midrule
Terra MODIS MOD09A1 Collection 6.1 & From  {18 February 2000;} %MDPI: Date should be Day Month Year format. Please check the whole table and revise them. @VP: done. All dates in the table now use Day Month Year, including 4 July 2015 for the reprocessed Sentinel-2 archive.
 8-day composite & 500 m & 1500  {$\times$} %MDPI: We revised the letter “x” to a multiplication sign (“×” U+00D7). Please confirm. Same below. @VP: done
 1500 m & b01/b02/b06 \\\midrule
Aqua MODIS MYD09A1 Collection 6.1 & From  {4 July 2002}; 8-day composite & 500 m & 1500  {$\times$} 1500 m & b01/b02/b06 \\\midrule
VIIRS VNP09H1 Collection 2 & From  {19 January 2012}; 8-day composite & 500 m & 1500  {$\times$} 1500 m & I1/I2/I3 \\\midrule
Landsat 8/9 OLI/OLI-2 Collection 2 Level-2 & From  {11 April 2013}; surface reflectance & 30 m & 90  {$\times$} 90 m & B4/B5/B6 \\\midrule
Sentinel-2 MSI Level-2A & From  {4 July 2015}; reprocessed surface reflectance & 20 m & 60  {$\times$} 60 m & B4/B8A/B11 \\\midrule
Sentinel-3 SY\_2\_SYN 300 m & From  {1 May 2018}; surface reflectance & 300 m & 900  {$\times$} 900 m & Oa08/Oa17/S5N \\
\bottomrule
\end{tabularx}
\end{adjustwidth}
\end{table}

Sentinel-2 and Landsat candidates must have scene-level cloud cover below
30\%; pixel masks then remove the specified cloud, shadow, snow and
invalid/saturated observations. MODIS and VIIRS use their state-quality
flags and require valid reflectances in the interval $(0,1.6)$.
Sentinel-3 SYNERGY includes cloud and snow screening during Level-2
processing. Cloudy, sea and invalid pixels are excluded from the 15-by-15
pixel averages used for aerosol retrieval \citep{copernicus2026s3synprocessing}.
Our sampler reads the resulting reflectances with NetCDF fill-value masking
and scale/offset decoding, and counts finite values without applying an
additional cloud, snow or aerosol quality-flag mask. Its valid-pixel fraction
therefore describes the available reflectances after product processing.
The exact masks, reflectance scaling and product-specific differences are
documented in Supplementary File S3.

All products retain central-pixel reflectance for the final LFMC predictors.
The output uses the product-native grid: 500 m for MODIS and VIIRS, 300 m for
Sentinel-3, 30 m for Landsat and 20 m for the Sentinel-2 common-band
implementation. A surrounding 3-by-3~native-pixel window is sampled for
land-cover representativeness and valid-pixel support screening, calibration
pairing and supplementary footprint sensitivity. These windows are 1500 by
1500 m for MODIS and VIIRS, 900 by 900 m for Sentinel-3, 90 by 90 m for
Landsat and 60 by 60 m for Sentinel-2. GlobeLFMC supplies point coordinates
without field polygons; the surrounding land-cover checks therefore assess
whether the central observation lies in a sufficiently vegetated and
compositionally consistent local~\mbox{environment}.

The sampler retains central values, footprint means, medians, standard
deviations, valid-pixel counts and fractions. Diagnostics also include
fractions with normalized difference vegetation index (NDVI) below 0.05 or
at least 0.20, and European Space Agency (ESA) WorldCover composition.
Only the central-pixel common-band indices enter the final model; footprint
summaries and diagnostics are used for screening, calibration and
supplementary analyses. Missing spectral placeholders are decoded before
calibration and index calculation, so missing central pixels remain excluded.

Sentinel-2 MSI Level-2A was sampled at 20 m because the final common optical
feature space requires the red, NIR and SWIR1 bands. We therefore used B8A
as the Sentinel-2 NIR band rather than the 10 m broad B8 band, so that the
red, narrow-NIR and SWIR-sensitive predictors could be constructed on the
native 20 m support of B8A and B11, and so that resampling the SWIR1 band to 10 m could be avoided.

\subsubsection{Meteorological, Topographic and Seasonal Predictors}

The empirical model is not a satellite-only model. GlobeLFMC 2.0 provides a
richer set of ancillary information than the original GlobeLFMC release,
including meteorological descriptors that can be matched to each field
measurement \citep{yebra2024globelfmc}. We exploit this improvement by
combining optical vegetation indices with weather, topography and seasonal
predictors. The weather variables include precipitation accumulated over
multiple antecedent windows, relative humidity, air temperature, vapour
pressure, wind speed, dewpoint and vapour-pressure deficit derived from
temperature and vapour pressure. These variables represent both short-term
drying and longer drought-related controls on plant water status~\mbox{\citep{munoz2021era5land, ruffault2018droughtindices}}. Topography is resampled
consistently from the Advanced Land Observing Satellite (ALOS) World 3D 30 m
(AW3D30) digital surface model and includes elevation, slope and aspect
\citep{tadono2014alos}. Day of year is encoded as sine and cosine to represent
seasonal cycles without a discontinuity between December and January.

The complete weather predictor set is listed in
Table~\ref{tab:weather-predictors}.

\begin{table}[H]
\caption{Meteorological predictors used by the compact unified feature set. Variables are inherited from the GlobeLFMC 2.0 ancillary data table and are matched to each field sampling date. Precipitation windows represent antecedent accumulated rainfall, while humidity, temperature, vapour-pressure and wind variables describe atmospheric demand and short-term drying conditions.}
\label{tab:weather-predictors}

\begin{adjustwidth}{-\extralength}{0cm}
%\centering %% If there is a figure in wide page, please release command \centering, for Table, ``\textwidth" should be ``\fulllength"
\begin{tabularx}{\fulllength}{@{\extracolsep{\fill}}>{\raggedright\arraybackslash}m{0.15\fulllength}>{\raggedright\arraybackslash}m{0.4\fulllength}>{\raggedright\arraybackslash}m{0.4\fulllength}@{}}
\toprule
\textbf{Predictor Group} & \textbf{Variables} & \textbf{Role in LFMC Retrieval} \\
\midrule
Antecedent rainfall & 24 h precipitation; 3-day precipitation sum; 1-week precipitation sum; 4-week precipitation sum; 12-week precipitation sum & Represents short-term wetting and progressively longer drought-memory controls on plant water status. \\\midrule
Relative humidity & 2 m relative humidity at 06:00, 09:00, 12:00 and~15:00 & Captures sub-daily atmospheric moisture conditions associated with evaporative demand. \\\midrule
Temperature & 2 m daily maximum air temperature; 2 m daily mean air temperature & Represents thermal forcing affecting drying rate and plant water stress. \\\midrule
Vapour-pressure terms & 24 h mean vapour pressure; 24 h mean dewpoint temperature; vapour-pressure deficit derived from air temperature and vapour pressure & Summarises atmospheric water demand and moisture availability near the canopy. \\\midrule
Wind & 10 m daily mean wind speed & Accounts for ventilation effects that can enhance drying under otherwise similar humidity and temperature conditions. \\
\bottomrule
\end{tabularx}
\end{adjustwidth}
\end{table}

Additional product-specific variable sets were evaluated before the final
compact predictor set was selected. These included Sentinel-1 radar
backscatter and incidence angle for high-resolution optical branches, Landsat
thermal surface-temperature predictors, and VIIRS VNP15A2H leaf area index (LAI) and fraction of absorbed
photosynthetically active radiation (FPAR) layers
\citep{torres2012sentinel1, masek2020landsat9,nasa_vnp15a2h_002}. These variables are described in
Section~\ref{sec:methods} because they were part of the feature-selection
process, but they are not retained in the final unified model. The promoted
feature set therefore reflects an explicit screening step: it keeps the
predictor families that improved LFMC retrieval consistently while avoiding
product-specific dependencies that would reduce the multisensor coverage.

\subsection{Processing and Modelling Workflow}
\label{sec:methods}

\subsubsection{Remote-Sensing Site Suitability}

GlobeLFMC sites are point locations, not mapped field polygons. This creates a
representativeness problem: a 1500 m MODIS or VIIRS 3-by-3 footprint may contain the
field vegetation, roads, bare soil, cropland, water or other vegetation types.
Without explicit polygons we cannot know the exact sampled field area. We can,
however, reject site/product combinations whose satellite footprint is clearly
unsuitable for remote-sensing LFMC \mbox{model development}.

For every product, site and functional vegetation class, the sampled footprint
is overlaid on ESA WorldCover 2021 at 10 m resolution
\citep{zanaga2022worldcover}. A site/product/vegetation combination is retained
using product-specific spatial-support rules. The central footprint refers to
the WorldCover cells falling inside the native satellite pixel centred on the
GlobeLFMC site, whereas the full footprint is the surrounding 3-by-3 native-pixel
support. For Landsat 8/9 Collection 2 Level-2 and Sentinel-2 MSI Level-2A, the
central footprint must contain at least 90\% natural vegetation; the full 3-by-3
footprint must contain at least 80\% natural vegetation, non-vegetated classes
must account for at most 15\% of the full footprint and at least 70\% of the full
footprint must match the dominant central land-cover class. For MODIS, VIIRS and
Sentinel-3 SY\_2\_SYN 300 m, the central-location criterion is kept strict but
the full-footprint criterion is relaxed to account for the larger spatial
support: the central footprint must contain at least 85\% natural vegetation,
the full footprint must contain at least 70\% natural vegetation, non-vegetated
classes must account for at most 25\% of the full footprint and at least 60\% of
the full footprint must match the dominant central class. These rule-based
thresholds were selected through a preliminary sensitivity analysis of five
profiles spanning central natural-vegetation requirements of 50--90\% and
progressively relaxed full-footprint requirements. The analysis considered the
retained site/product/class combinations, retained field matches and preliminary
within-site model behaviour, after which the selected profile was frozen before
the final training run. Under the selected profile, 69--82\% of high-resolution,
66--78\% of Sentinel-3 and 61--75\% of MODIS/VIIRS source rows are retained,
depending on vegetation class. These raw-source percentages precede LFMC
adequacy and central-predictor completeness filtering. The final thresholds
preserve a strict local-site
requirement while avoiding unrealistic purity demands over moderate-resolution
footprints. The criteria are evaluated separately for each
product support. A site can
therefore be acceptable for Landsat or Sentinel-2 but rejected for MODIS if the larger MODIS
footprint mixes the field vegetation with non-target surfaces.
These criteria select field-matched training records. The mapping example in
Appendix~\ref{app:camp-fire-case} uses land cover to select a vegetation-class
model and central-pixel quality screening, without applying these training
suitability criteria as inference masks.

Valid-pixel support is a separate observation-level quality check. The
WorldCover analysis above is a one-off site/product suitability assessment
based on land-cover composition. The valid-pixel filter is applied to every
sampled satellite observation before training, using the product quality masks and the
actual valid pixels available on that date. Each satellite sample must contain at least
three valid pixels. Landsat 8/9 Collection 2 Level-2 and Sentinel-2 MSI Level-2A rows must
have at least 75\% valid-pixel fraction, while MODIS, VIIRS and Sentinel-3 rows must have
at least 60\%. At the site/product level, the median valid-pixel fraction must
also meet the same product-specific threshold. These filters remove cases
where a satellite observation is dominated by missing data, cloud/snow
masking, footprint-edge artefacts or other non-representative support.

These filters permit product-specific footprint sizes by retaining sites whose
surrounding land-cover support is sufficiently homogeneous at the corresponding
spatial resolution. This reduces, but cannot eliminate, representativeness error
for coarse-resolution products: without field-site polygons, homogeneity around
the reported coordinate is only a proxy for agreement between the sampled LFMC
and the vegetation represented by the satellite footprint. Computing pixel
statistics over a 3-by-3 footprint was evaluated as a sensitivity analysis
(Supplementary File S5). The final predictors use central pixels; neighbourhood
composition remains a suitability criterion. Field-coordinate uncertainty is
not explicitly quantified in GlobeLFMC 2.0.

\subsubsection{LFMC Time-Series Adequacy}

Site suitability for remote sensing does not necessarily imply site suitability for
model validation. A site may be spectrally representative but still uninformative if it
has too few LFMC observations or if all observations fall in a narrow seasonal
window. This can happen because monitoring campaigns differ in sampling
frequency, duration and seasonality, and because some species or site
conditions have intrinsically flat LFMC curves. We therefore apply the time-series adequacy filter to the aggregated
site-date-functional-class target series only after product-specific site
suitability, observation-level satellite quality control and completeness of
the predictor set selected for the corresponding experiment have been
enforced. A site/class series must contain at least 20 field dates, span at least 365
days and show sufficient LFMC dynamic range. The minimum site-level LFMC
standard deviation is 20 percentage points for Grass and Shrub and 30
percentage points for Tree.

Time-series adequacy defines the field-record population used to evaluate
retrieval of LFMC dynamics. Observation count, temporal span and LFMC standard
deviation are calculated from the complete eligible site/class series after
satellite quality, site-suitability and selected-predictor completeness
screening. This record-level assessment retains sufficiently sampled,
multi-season series with appreciable LFMC variability and avoids rejecting a
series because an individual training fold covers a shorter or seasonally
restricted subset. The dynamic-range criterion does not imply that stable LFMC
records are erroneous; it restricts the population considered in this analysis.

Screening field observations before retrieval validation has precedent in
Quan et~al.~\citep{quan2021}, who applied a temporal outlier rule based on
consecutive FMC observations and the time-series standard deviation, together
with a spatial homogeneity filter. Their temporal rule addresses abrupt
changes in individual measurements; the present criterion addresses the
dynamic range of an entire site/class record.

The complete-series assessment includes the LFMC values subsequently held out
for evaluation. Those values contribute to cohort eligibility but are excluded
from model fitting and from estimation of fold-specific preprocessing
parameters. Accordingly, the reported cross-validation performance is
conditional on the retrospectively selected, sufficiently variable records;
it does not establish performance for records excluded by the adequacy
criteria. The numerical adequacy thresholds are held constant across the final
model comparisons, and their effect on sample support is reported explicitly.

The one-year span requirement is important. A low LFMC standard deviation can
mean that a site is genuinely stable, but it can also mean that the site was
sampled only during a short part of the annual cycle. Requiring both a minimum
number of dates and a minimum temporal span turns the standard-deviation
threshold into a meaningful dynamic-range criterion while preventing short
records from being rejected simply because they cover only part of the annual
cycle.

Table~\ref{tab:unified-final-support} reports the final all-products support
after LFMC range filtering, time-series adequacy, WorldCover footprint
suitability and valid-pixel support. MODIS contributes the largest number of
retained events for each vegetation class because it covers the longest period.
Aqua support is additionally limited by missing SWIR1 reflectances, as discussed
for Figure~\ref{fig:unified-year-counts}, and by the valid-pixel support requirement.
This, and the fact that its historical record now allows for historical timeseries
analyses that span more than two decades, are the main motivations for including MODIS in the unified model, despite
the platform age and added complexity in dataset handling and curation associated with
the increased size.

\begin{table}[H]
\caption{Product rows and sites retained by the final central-pixel all-products models after field screening, LFMC ranges, neighbourhood-based WorldCover and valid-pixel suitability, complete central-pixel predictors and LFMC time-series adequacy. Each evaluated field event receives equal total metric weight across its available products.}
\label{tab:unified-final-support}
\begin{tabularx}{\textwidth}{lLRR}
\toprule
\textbf{Product} & \textbf{Vegetation} & \textbf{Product Rows} & \textbf{Sites} \\
\midrule
Landsat 8/9 C2 L2 & Grass & 1201 & 58 \\
Landsat 8/9 C2 L2 & Shrub & 7771 & 210 \\
Landsat 8/9 C2 L2 & Tree & 3870 & 97 \\
Terra MODIS & Grass & 5259 & 80 \\
Terra MODIS & Shrub & 22,212 & 228 \\
Terra MODIS & Tree & 14,216 & 110 \\
Aqua MODIS & Grass & 1777 & 80 \\
Aqua MODIS & Shrub & 8484 & 227 \\
Aqua MODIS & Tree & 5065 & 110 \\
Sentinel-2 MSI L2A & Grass & 1479 & 46 \\
Sentinel-2 MSI L2A & Shrub & 8523 & 213 \\
Sentinel-2 MSI L2A & Tree & 4091 & 98 \\
Sentinel-3 SYN 300 m & Grass & 875 & 37 \\
Sentinel-3 SYN 300 m & Shrub & 5937 & 198 \\
Sentinel-3 SYN 300 m & Tree & 2768 & 90 \\
VIIRS VNP09H1 & Grass & 2915 & 58 \\
VIIRS VNP09H1 & Shrub & 14,510 & 220 \\
VIIRS VNP09H1 & Tree & 7966 & 104 \\
\bottomrule
\end{tabularx}
\end{table}

\subsubsection{Common Spectral Feature Engineering}

The final unified model uses a strict common optical feature space. Red, NIR
and SWIR1 are available across Terra MODIS MOD09A1, Aqua MODIS MYD09A1, VIIRS
VNP09H1, Landsat 8/9 Collection 2 Level-2, Sentinel-2 MSI Level-2A and
Sentinel-3 SY\_2\_SYN 300 m, and
they support the
vegetation indices in Table~\ref{tab:common-vi}.
Across the supported instruments, these bands are centred approximately at
0.65, 0.86 and 1.6~$\upmu$m, respectively. They cover corresponding spectral
regions, but their effective wavelengths, bandwidths and response shapes differ.
We account for these differences using the instrument-specific SRFs from the
RTTOV/NWP SAF library \citep{nwpsaf2026rttov_srf}; their compatibility and the
reflectance calibration are evaluated below.
The final common optical feature set comprises red, NIR and SWIR1, the spectral
regions available in every retained product. This selection allows the
framework to make full use of the MODIS MOD09A1/MYD09A1 and VIIRS VNP09H1
records. Their quality-screened 8-day, 500 m composites provide regular
observations that are convenient for historical analyses and repeated LFMC
mapping at national to continental scales through Google Earth Engine. MODIS
supplies the longest satellite record, while VIIRS extends moderate-resolution
coverage into the current operational period. Blue, green, red-edge and SWIR2
bands are excluded because they are unavailable in one or more of these
products. Landsat and Sentinel-2 nevertheless retain their finer spatial
support for local applications.

The selected indices provide complementary information on vegetation
condition. The normalized difference vegetation index (NDVI) and two-band
enhanced vegetation index (EVI2) respond to greenness and canopy cover. The
normalized difference infrared index (NDII), moisture stress index (MSI) and
global vegetation moisture index (GVMI) combine NIR and SWIR1 reflectance to
provide information on vegetation water content~\mbox{\citep{tucker1979, hardisky1983, hunt1989, ceccato2002, jiang2008evi2,
yebra2013review}}.

\begin{table}[H]
\caption{Common vegetation indices used by the final unified model. NDVI is the Normalized Difference Vegetation Index, NDII is the Normalized Difference Infrared Index, MSI is the Moisture Stress Index, GVMI is the Global Vegetation Moisture Index and EVI2 is the two-band Enhanced Vegetation Index. The same formulas are computed after empirical reflectance calibration for each tested pixel statistic.}
\label{tab:common-vi}
\renewcommand{\arraystretch}{1.2}
\begin{tabularx}{\textwidth}{llLl}
\toprule
\textbf{Index} & \textbf{Formula} & \textbf{Main Interpretation} & \textbf{Reference} \\
\midrule
NDVI & $\frac{NIR-Red}{NIR+Red}$ & Canopy greenness and cover & \citep{tucker1979} \\[1ex]
NDII & $\frac{NIR-SWIR1}{NIR+SWIR1}$ & Canopy water-content sensitivity & \citep{hardisky1983} \\
MSI & $\frac{SWIR1}{NIR}$ & Moisture stress; higher values indicate drier canopies & \citep{hunt1989} \\
GVMI & $\frac{(NIR+0.1)-(SWIR1+0.02)}{(NIR+0.1)+(SWIR1+0.02)}$ & Global vegetation moisture index & \citep{ceccato2002} \\
EVI2 & $\frac{2.5(NIR-Red)}{NIR+2.4Red+1}$ & Two-band enhanced vegetation index where blue is unavailable or inconsistent & \citep{jiang2008evi2} \\
\bottomrule
\end{tabularx}
\end{table}

For the satellite reflectance observations, "core" %MDPI: Please confirm if the italics are necessary; if not, please remove them. The following highlights are the same. @VP: replaced with quotes, implying that I am defining the terms here. The remaining highlighted scenario labels have also been changed to quotation marks.
 denotes the reflectance
of the central pixel sampled at the field coordinate. The "mean" and
"median" summaries are calculated from valid reflectances in the surrounding
3-by-3 sampling window at the product's native grid resolution. These spatial
summaries are distinct from the arithmetic-mean LFMC target calculated from
field measurements at the same site, date and functional class. After empirical
reflectance calibration, the indices are recomputed from each reflectance
summary so that the predictors correspond to the tested spatial sampling method.
Central-pixel reflectance is used in the final models; 3-by-3 mean and median
summaries are retained in the sensitivity analyses.
%MDPI: Table citation order is wrong. Table 9 should be cited after Table 8. Please check and revise the citations to make sure the first citation of each table appears in numerical order. @VP: Removed the premature Table 9 citation and corrected the first-citation order of appendix tables. Table positions and numbering are unchanged.

\begin{figure}[H]
\includegraphics[width=\textwidth]{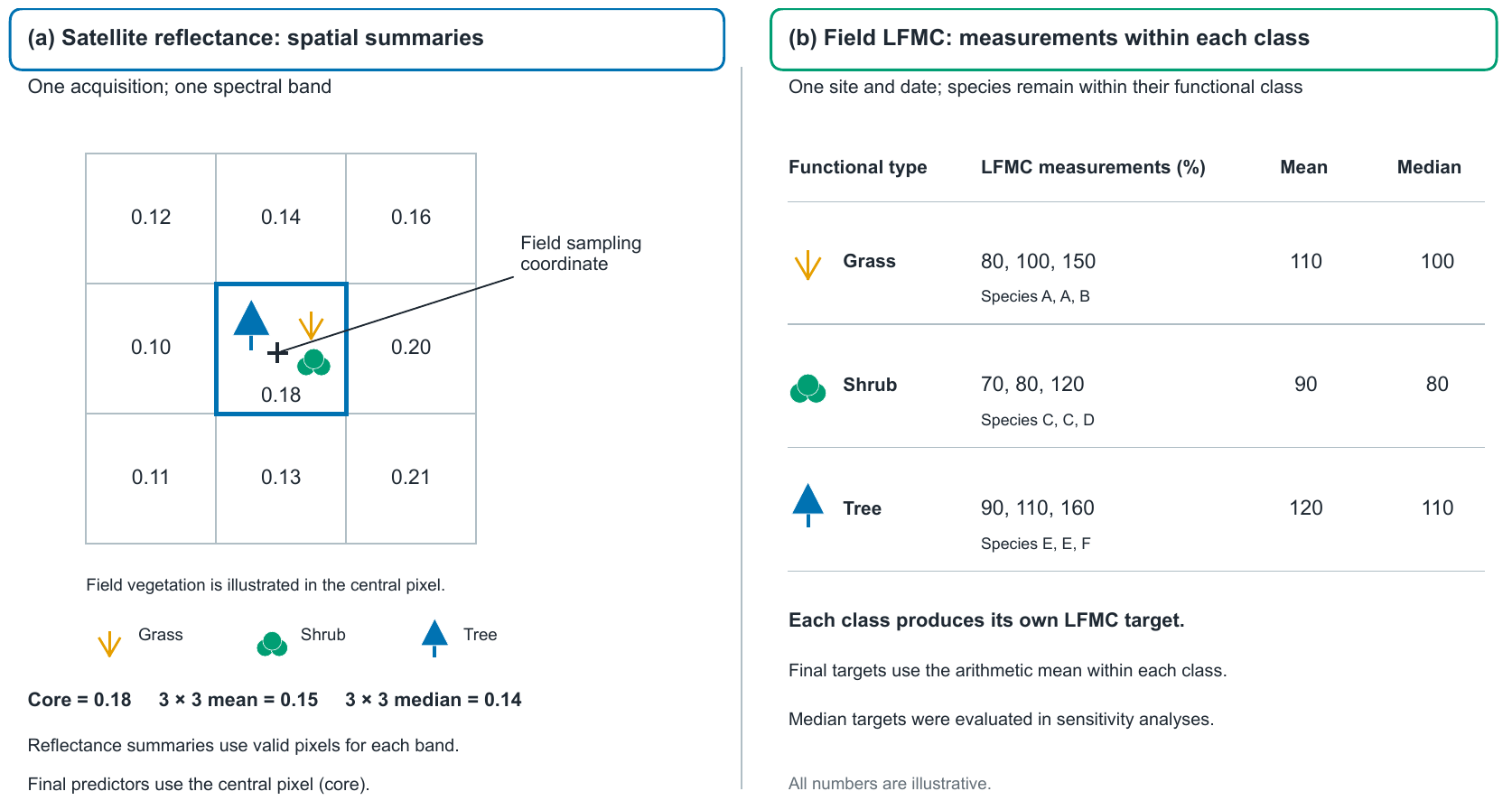}
\caption{Satellite reflectance summaries and field LFMC targets. (\textbf{a}) The core value is the reflectance of the pixel containing the field sampling coordinate. Spatial mean and median reflectances are calculated separately for each spectral band from valid pixels in the surrounding 3 $\times$ 3 window, including the central pixel. Vegetation symbols illustrate the presence of different functional types. (\textbf{b}) Field LFMC measurements from the same site and date are grouped by functional class. Each class has its own target, calculated from its measurements, which may include more than one species. The final models use central-pixel reflectances and arithmetic-mean LFMC targets; alternative spatial summaries and median LFMC targets are evaluated in sensitivity analyses. All values are illustrative.}
\label{fig:pixel-field-aggregation}
\end{figure}

Figure~\ref{fig:pixel-field-aggregation} illustrates the distinction between
spatial reflectance summaries and LFMC targets calculated from field
measurements within each functional class.

\subsubsection{Preliminary Feature-Set Screening}

Before fixing the unified multisensor architecture, we tested whether
product-specific optional variables materially improved LFMC retrieval under
the primary within-site date-holdout validation. In this validation mode,
observations are split so that each retained site can contribute records to
both training and testing, but the held-out records correspond to dates not
used to fit the model. The test therefore evaluates whether a model trained on
the multisite archive can estimate additional dates at sites whose broad
ecological conditions are already represented in training. 

The preliminary feature-set screening was deliberately performed on representative
single-product branches, because optional variables are not available for all
products and would otherwise complicate the sensor-unification step. The
high-resolution screening used Sentinel-2 MSI Level-2A and Landsat 8/9
Collection 2 Level-2. The moderate-resolution screening used VIIRS VNP09H1.
These preliminary development experiments used the same LFMC bounds,
site-level criteria, mean target aggregation, candidate model families and
primary validation design. Their tested footprint summaries are retained as
historical feature-screening results; the revised production models use
central-pixel predictors~\mbox{consistently}.

The tested feature sets were ordered from minimal to progressively richer
predictor configurations. First, models were trained with optical vegetation
indices alone. Second, meteorological variables were added. Third, topography
and cyclic day-of-year were added, forming the compact candidate feature set.
The optional product-specific tests then added Sentinel-1 vertically transmitted, vertically received (VV) and
vertically transmitted, horizontally received (VH) backscatter channels and incidence-angle features to the
high-resolution branches, Landsat thermal surface-temperature features to the
Landsat branch, and VIIRS VNP15A2H leaf area index/fraction of absorbed
photosynthetically active radiation (LAI/FPAR) features to the VIIRS branch.
This experiment was used to select the predictor
family adopted by the final multisensor model; the model itself is
trained and evaluated separately in the unified
MODIS/VIIRS/Landsat/Sentinel-2/Sentinel-3 branch.

\subsubsection{SRF Diagnostics and Empirical Multisensor Calibration}

Pooling satellite products requires explicit treatment of differences in
spectral response, spatial support and compositing. The unified branch
therefore uses two complementary harmonisation steps. First, SRFs from the
radiative-transfer software database maintained by the Numerical Weather
Prediction Satellite Application Facility (NWP SAF) are standardised into a common table
and used to quantify the compatibility of Terra MODIS, Aqua MODIS, VIIRS,
Landsat 8/9, Sentinel-3 and Sentinel-2 red, NIR and SWIR1 bands~\citep{nwpsaf2026rttov_srf}. For each
common band, we compute effective wavelength, full width at half maximum
(FWHM) and a normalised spectral overlap index relative to the
Sentinel-2 reference domain. Effective wavelength, $\lambda_{\mathrm{eff}}$, is the response-weighted mean  {wavelength,} %MDPI: Please ensure all variables/values in the equation appear in the same format in the text (normal/italic/bold/subscript/superscript). @VP: Added explicit mathematical symbol definitions and used consistent italic variables and upright descriptive subscripts in the text and equations.
\[
\lambda_{\mathrm{eff}}=\frac{\int \lambda R(\lambda)\,d\lambda}{\int R(\lambda)\,d\lambda},
\]
where $\lambda$ is wavelength and $R(\lambda)$ is the spectral response
function. Wavelength differences relative to Sentinel-2 are denoted by
$\Delta\lambda_{\mathrm{eff}}$. The overlap index, $O$, is computed after
normalising each SRF to unit area as
\[
O=\int \min\left(R_1'(\lambda),R_2'(\lambda)\right)\,d\lambda,
\]
where $R_1'(\lambda)$ and $R_2'(\lambda)$ are the unit-area-normalised
response functions of the product band and the Sentinel-2 reference band,
respectively. Therefore, $O=1$ indicates identical normalised responses and
lower values indicate lower spectral compatibility with the Sentinel-2
reference band.

Second, paired acquisitions are used to learn target-independent reflectance
transformations. Candidate pairs must refer to the same GlobeLFMC site and field-observation
date, and the actual acquisition
dates represented by the recorded product timestamps must differ by no more
than one day. For MODIS and VIIRS, these are composite dates, so this criterion
does not establish one-day agreement between the underlying pixel acquisitions.
If both Sentinel-2A and Sentinel-2B acquisitions satisfy this condition,
the temporally closest observation is selected using deterministic
tie-breaking. Sentinel-2 MSI Level-2A is selected as the reference domain
because it provides the finest spatial resolution among the retained
common-band implementations: B4, B8A and B11 are represented on the 20~m
B8A/B11 grid. The two-satellite Sentinel-2 constellation also has a shorter
nominal revisit interval than the combined Landsat 8/9 constellation. These
characteristics provide spatially detailed reference observations and frequent
temporal overlap with the other satellite records used to estimate the
empirical reflectance transformations \citep{drusch2012sentinel2,masek2020landsat9}.

A separate Huber-regression transformation is estimated for each instrument
and common band, thereby distinguishing, for example, Landsat 8 from Landsat 9,
Sentinel-3A from Sentinel-3B, and Terra MODIS from Aqua MODIS
\citep{huber1964robust}. Each fit uses mean reflectance and requires at least 100
valid acquisition pairs. SRFs establish whether the selected bands represent
compatible spectral regions and motivate calibration, but do not enter the
regression as predictors. The empirical fits can therefore account jointly for
first-order SRF differences and systematic effects associated with atmospheric
correction, compositing, acquisition geometry and spatial support.

The production band mapping is fitted on footprint-mean pairs and applied
to the central-pixel reflectances used by the LFMC models. Footprint means
provide calibration pairs even when an individual central observation is
missing; missing central values are not imputed for LFMC prediction.
Sensitivity to this choice is evaluated by fitting core-specific mappings
against central-pixel Sentinel-2 reflectance within the same acquisition-safe
outer calibration pools. Mean-derived and core-specific mappings are compared
on identical held-out core pairs, including production clipping, with residual
summaries by available IGBP land-cover code and calendar quarter. All six
central-pixel LFMC configurations are refitted with the alternative mappings
while targets, rows, folds, model settings and weights remain fixed.
Supplementary File S6 reports diagnostics, pair counts, fold coefficients and
paired LFMC difference intervals.

Calibration is cross-fitted with the LFMC evaluation. For each outer validation
fold, the calibration corpus comprises all eligible, site-suitable paired
reflectances after excluding every connected component containing a held-out
field site/date or reused acquisition. Connections include the additional
acquisition identifiers retained for composite records. The calibration is independent of vegetation class and uses no
LFMC values. Its coefficients are then applied to the central-pixel
red, NIR and SWIR1 reflectances in both the LFMC training and test partitions,
with adjusted values clipped to [$-$0.05,1.50], after which all common
vegetation indices are recomputed. Calibration
coefficients are fitted once from the complete eligible reflectance archive and
stored with the final production model only after cross-validated evaluation is
complete. This procedure avoids direct use of test-fold spectra when estimating
the fold-specific transformations, while retaining enough overlap observations
to estimate stable instrument-level relationships. The resulting conversion is
empirical and is therefore applicable within the spectral and environmental
domain represented by the paired acquisitions rather than as a universal
radiometric transformation.

Two product scenarios are evaluated. The `` {all-products}'' scenario includes
Terra MODIS MOD09A1, Aqua MODIS MYD09A1, VIIRS VNP09H1, Landsat 8/9
Collection 2 Level-2, Sentinel-2 MSI Level-2A and Sentinel-3 SY\_2\_SYN
300 m. The `` {no-MODIS}''
scenario excludes both MODIS products and retains VIIRS, Landsat 8/9
Collection 2 Level-2, Sentinel-2 MSI Level-2A and Sentinel-3 SY\_2\_SYN
300 m. MODIS extends the archive back to 2000 and increases the number of
retained field events. The two scenarios have different eligible populations
and provide descriptive sample-support and performance summaries. The separate
controlled ablation isolates the contribution of MODIS training observations
conditional on shared fold-local reflectance calibration.

\subsubsection{Machine-Learning Models}
\label{sec:ml-models}

\textls[-15]{Two model families are retained for the final unified experiments: Random
Forest~\citep{breiman2001randomforest} and XGBoost
\citep{chen2016xgboost}.~Both are tree-ensemble methods suited to tabular
remote-sensing regression because they can represent nonlinear responses,
interactions among spectral and environmental predictors, and mixed-scale
variables without requiring the predictors to be linearly related to LFMC.
Random Forest trains many regression trees on bootstrap samples and averages
their predictions, which provides a stable nonlinear ensemble baseline and has
already been used successfully in LFMC mapping \citep{cunill2022rf}. XGBoost
builds an additive sequence of regression trees, where each new tree is fitted
to reduce the residual error of the current ensemble while using
regularisation and shrinkage to limit overfitting. During preliminary
development we also tested support-vector regression, multilayer perceptron
regressors and HistGradientBoosting using the  {scikit-learn Python library (version 1.9.0)} %MDPI: Please state which version of the software was used. @VP: Added scikit-learn version 1.9.0, verified against our current environment and its conda installation history.
~\citep{pedregosa2011scikit}.} These alternatives were not retained in the final grid because they consistently underperformed compared with Random Forest and XGBoost in the screening runs or introduced additional tuning complexity without a corresponding performance gain.

Mean LFMC and central-pixel reflectance were fixed during revision for all
classes and scenarios. Random Forest is retained for the all-products models
and for no-MODIS Shrub and Tree; XGBoost is retained for no-MODIS Grass,
following the original central-pixel development comparisons. These model
families were selected using development cross-validation, including the
reported folds. The corrected fixed-configuration results therefore remain
conditional on that development selection.

Table~\ref{tab:model-hyperparameters} reports the main settings used in the
final experiment grid. The complete estimator configuration, software versions,
feature list and filtering thresholds are also written automatically to each
run manifest so that the exported models can be traced to the exact training
configuration.

\begin{table}[H]
\caption{Main hyperparameters used for the evaluated regression models. Random Forest was selected for the final Grass, Shrub and Tree models; XGBoost was retained as the principal alternative candidate. Parameters not listed here retain the library defaults recorded in the machine-readable run manifest.}
\label{tab:model-hyperparameters}
\begin{tabularx}{\textwidth}{lL}
\toprule
\textbf{Model} & \textbf{Hyperparameter Settings} \\
\midrule
Random Forest & n\_estimators = 200; criterion = squared\_error; max\_depth = None; max\_features = sqrt; min\_samples\_split = 2; min\_samples\_leaf = 2; bootstrap = True; max\_samples = 0.9; random\_state = 42 \\\midrule
XGBoost & n\_estimators = 500; learning\_rate = 0.03; max\_depth = 4; \mbox{subsample = 0.8}; colsample\_bytree = 0.8; tree\_method = hist; random\_state = 42 \\
\bottomrule
\end{tabularx}
\end{table}

Models are trained separately for Grass, Shrub and Tree. The final
sensor-agnostic branch does not include product identity, product family or
footprint size as predictors. Those variables are used only for filtering,
auditing and validation. This choice is intentional: if product identity were
included, a high score could reflect product-specific calibration shortcuts
rather than a common LFMC retrieval model.

Before each model fit, redundant predictors are removed using a fold-specific
correlation-pruning step. Pearson correlations are computed from the training
predictor matrix only, after the complete-feature filter and before model
fitting. When two candidate predictors have an absolute correlation greater
than 0.95, the later predictor in the feature list is removed for that fold;
the same rule is then applied during the final refit on the full training set.
This reduces near-duplicate information among vegetation indices and ancillary
variables without using held-out observations, and has the added benefit of
simplifying the variables necessary to run the models at inference. The sine and cosine day-of-year
terms are exempt from this pruning rule so that the cyclic seasonal encoding
remains two-dimensional.

Two forms of unequal representation are addressed during model fitting. First,
the number of available satellite products differs among field events. If an
event $e$ is represented by $n_e$ products, each corresponding product row is
assigned the base weight $1/n_e$. The rows associated with an event therefore
have a combined weight of one, irrespective of satellite availability, while
remaining available as distinct spectral observations for multisensor~\mbox{training}.

Second, LFMC values are unevenly distributed, with many events concentrated in
intermediate moisture ranges and fewer events in very dry or very wet
conditions. Within each training fold, events are grouped into 10
percentage-point LFMC bins. Bin frequencies are calculated from the sum of the
event-balanced row weights; consequently, a multiply observed field event is
counted once rather than once per available product. For a row $i$ belonging
to event $e$ and LFMC bin $b$, the fitting weight is proportional to
\begin{equation}
w_i = \frac{1/n_e}{N_b},
\end{equation}
where $N_b$ is the event-equivalent mass in bin $b$. The weights are normalised
to a mean of one and passed to Random Forest and XGBoost through their
observation-weight interfaces. This combined weighting prevents both
well-observed events and densely populated LFMC intervals from dominating the
fitting loss.

The target distribution is also transformed before fitting because LFMC values
are concentrated in some moisture ranges and sparse in others. This step must
account for the fact that one field event can produce several product rows. For
example, if the same site--date LFMC measurement is matched to MODIS, VIIRS and
Sentinel-2, the training table contains three spectral observations but only one
independent LFMC target. Using all three rows to estimate the target distribution
would count that LFMC value three times and would give events observed by more
products disproportionate influence over the transformation. We therefore
retain one target value for each unique site--date--functional-class event when
fitting the quantile transformation. The observed training-event LFMC values are
ranked and their empirical quantiles are mapped to the corresponding quantiles
of a normal distribution. The resulting mapping is then applied to the target
of every training product row. Models are fitted in this transformed target
space, and their predictions are converted back to LFMC percentage units before
metrics are calculated. The mapping is estimated independently within each
training fold; LFMC values assigned to the corresponding test fold do not
contribute to it. Event counts and LFMC-bin frequencies used for observation
weighting are likewise calculated from the training fold only. The sole
full-series criterion applied before splitting is the predeclared time-series
adequacy filter described above.
\subsubsection{Validation Design}
\label{sec:validation-design}

The primary validation mode is a within-site date-holdout experiment, which
evaluates within-site temporal interpolation. The five folds are assigned only
after rows that must not be separated have been grouped into validation blocks.
Two linking rules define these blocks. First, all satellite-product rows matched
to the same site--date--functional-class field event are linked because they
share one measured LFMC target. Second, rows from the same site and satellite
product are linked when they use at least one identical source acquisition. The
second rule is needed because an 8- or 10-day composite, or a temporally matched
scene, can be associated with more than one nearby field date. Placing one use
of that acquisition in training and another in testing would expose the model
to the same satellite observation on both sides of the validation split.

Links are propagated until every connected set of rows forms one block. For
example, if field events A and B share a MODIS composite, and event B and event
C share a VIIRS composite, A, B and C are assigned to the same block even if A
and C do not directly share an acquisition. Every row in a block is then placed
in the same fold. Blocks are assigned with five-fold stratification by site, and
a site contributes held-out predictions only when it contains at least five
independent blocks. Consequently, each evaluated site is represented in both
training and testing, while neither an LFMC field event nor a reused source
acquisition can occur on both sides of a fold. Test rows therefore represent
unseen field dates and source acquisitions at sites whose broader ecological
conditions are represented in training. This allows us to test our validation
design methodology: can a model trained on the multisite archive
estimate LFMC for additional observation dates at represented sites?

The primary temporal design keeps every evaluated site represented in training,
maximising use of the available field archive to evaluate additional dates at
represented sites. A complementary five-fold site-blocked diagnostic excludes
entire sites from LFMC fitting and reflectance calibration
(Appendix~\ref{app:site-blocked-validation};
Table~\ref{tab:site-blocked-validation}). It quantifies transfer to unseen
site identifiers within the archive's sampled domains. Neither validation mode
establishes global applicability or transfer to unsampled regions and biomes.

Complementary diagnostics isolate the contribution of MODIS training observations
and compare product-specific with multisensor training on matched held-out
records. Their experimental designs and complete tables are presented in
Appendix~\ref{app:multisensor-diagnostics}.

A controlled footprint sensitivity evaluates central-pixel, 3-by-3 mean and
3-by-3~median predictors using Random Forest and mean LFMC in the all-products
scenario. Rows complete under all three summaries are intersected, the
existing complete-series adequacy criteria are rechecked, and event weights
are recomputed. Targets, model settings, folds and mean-derived calibration
are identical across alternatives; pruning and transformations are fitted
within each training fold. Supplementary File S5 reports the matched metrics,
site-bootstrap intervals and support. These summaries remain supplementary
comparisons and are not candidate branches in the revised production grid.

A fixed-configuration sensitivity analysis evaluates stricter and more permissive
WorldCover profiles and removal of the minimum LFMC variability requirement.
The production filtering criteria are retained. Appendix~\ref{app:filter-sensitivity}
describes the alternative profiles, cohort reconstruction and results.

A paired environmental-only diagnostic quantifies the contribution of the
optical predictors. For each vegetation class, the promoted Random Forest
configuration is compared with a model using only weather, topography and
cyclic day-of-year. Both alternatives use identical retained product rows,
field-event targets, acquisition-blocked folds, training weights and
hyperparameters. Predictor correlation pruning, scaling and target
transformation are fitted within each training fold. The comparison is
conditional on the same satellite-matched and filtered population used by
the full model. Paired uncertainty follows the site-bootstrap procedure
described below.

Performance is reported using pooled \Rtwo, root mean square error (RMSE), Kling--Gupta efficiency (KGE),
Pearson correlation and bias. Pooled metrics use
the event weights $1/n_e$, so every held-out field event contributes the same
total mass regardless of the number of available satellite products. Individual
product predictions are retained in the output tables for product-specific
diagnostics; they are not averaged into a multisensor ensemble before the
headline metrics are calculated. KGE complements
\Rtwo by identifying models that reproduce correlation but distort the mean
or seasonal LFMC amplitude \citep{gupta2009kge}.

Uncertainty in the pooled multisensor metrics and the paired product-wise
comparisons is estimated using 10,000 bootstrap resamples of whole sites with
replacement (random seed 42). Each selected site retains all its held-out
observations, dates and evaluation weights, and the same site multiplicities
are applied to both models in each paired comparison. Weighted \Rtwo and RMSE
are recalculated for every resample; the 2.5th and 97.5th percentiles define
pointwise 95\% intervals. These intervals describe sampling uncertainty
conditional on the selected records, configurations and fitted out-of-fold
models. They do not incorporate model-selection uncertainty and are not
simultaneous intervals across the 18 comparisons. Complete metric intervals,
paired differences and support counts are provided in Supplementary File S4.

\section{Results}
\label{sec:results}

\subsection{Feature-Set Screening}

Table~\ref{tab:feature-ablation-gapfill} reports the preliminary
feature-set screening. The result is consistent across product families:
vegetation indices alone are not sufficient, adding weather improves the
models, and the compact set composed of vegetation indices, weather,
topography and cyclic day-of-year gives the major performance gain. For
Sentinel-2 MSI Level-2A, the compact set increases \Rtwo from the 0.458 obtained with vegetation indices only to 0.672
for Grass, from 0.453 to 0.675 for Shrub and from 0.422 to 0.663 for Tree
relative to vegetation indices alone. VIIRS VNP09H1 shows an even stronger
dependence on ancillary predictors: \Rtwo rises from 0.281, 0.171 and 0.120
with vegetation indices alone to 0.667, 0.687 and 0.599 with the compact set.

The optional variables do not provide enough additional skill to justify their
use in the final multisensor architecture. Sentinel-1 slightly reduces
Sentinel-2 performance and reduces the available sample size due to the requirement
of having observations from both satellites within a certain time window of the
GlobeLFMC field measurement. In the Landsat
branch, thermal information gives only a small Grass improvement
($\Delta\Rtwo=0.006$) and no meaningful Shrub or Tree gain.~VIIRS LAI/FPAR
features provide only small gains while reducing sample availability because
they require complete VIIRS VNP15A2H support. We therefore retain the compact
feature set for the final unified model: it captures most of the available
LFMC signal while remaining compatible with MODIS, VIIRS, Landsat,
Sentinel-2 MSI Level-2A and Sentinel-3 SY\_2\_SYN 300 m, and makes the
inference workflow \mbox{relatively agile}.

The environmental-only comparison (Table~\ref{tab:environmental-baseline}) gives
\Rtwo values of 0.700, 0.676 and 0.670
for Grass, Shrub and Tree, respectively. Adding central-pixel optical indices
changes \Rtwo by 0.015, 0.017, 0.021 and RMSE by
$-$1.13, $-$0.63, $-$0.86 LFMC percentage points. Paired intervals are reported
in the table; the substantial environmental baseline skill confirms the
predictive value of weather, topography and seasonal context.

\subsection{SRF and Overlap-Calibration Diagnostics}

Table~\ref{tab:srf-overlap-calibration} reports the SRF diagnostics and
empirical Huber-regression calibration fits for the non-reference products in the
all-products scenario. The SRF overlap confirms that the common bands are not
identical across products. MODIS has the weakest overlap with the
Sentinel-2 reference in SWIR1 and noticeable wavelength differences in red and
NIR. Sentinel-3 is closely aligned with Sentinel-2 in NIR and SWIR1 but has a
lower red-band overlap, while VIIRS and Landsat also retain product-specific
differences that justify explicit calibration. The robust linear
calibration reduces held-out central-pixel reflectance RMSE in 21 of 21 sensor/band combinations. The reductions
are modest, which is expected: the goal is not to erase all product
differences, but to remove systematic first-order offsets before recomputing
the common vegetation indices.

The central-pixel calibration sensitivity is reported in Supplementary File
S6 for all six fixed configurations. Relative to the retained mean-derived
mapping, core-specific mappings change \Rtwo from
$-$0.0034 to $-$0.0001 and RMSE from
+0.01 to +0.14 LFMC percentage points. All six point estimates favour the mean-derived mapping; both paired
intervals favour it for all-products Shrub and no-MODIS Tree, while the other
intervals include zero. The shared mean-derived production mapping is
retained; paired intervals and land-cover/quarter
reflectance diagnostics are provided in Supplementary File S6.

\begin{table}[H]
\caption{Historical development screening preceding the fixed central-pixel production configuration. Preliminary feature-set screening with mean LFMC fixed as the target under the primary within-site date-holdout validation. Compact denotes optical vegetation indices (VIs), weather, topography and cyclic day-of-year. LAI/FPAR denotes leaf area index and fraction of absorbed photosynthetically active radiation. Rows report the best mean-target model/statistic within each vegetation class and feature set.\label{tab:feature-ablation-gapfill}}
\small
\resizebox{\textwidth}{!}{
\begin{tabular}{lllllrr}
\toprule
\textbf{Experiment} & \textbf{Veg.} & \textbf{Model} & \boldmath{$n$} & \textbf{Stat.} & \boldmath{\Rtwo} & \textbf{RMSE} \\
\midrule
S2 VIs & Grass & RandomForest & 1967 & mean & 0.458 & 44.6 \\
S2 VIs & Shrub & RandomForest & 9565 & mean & 0.453 & 28.9 \\
S2 VIs & Tree & RandomForest & 6141 & mean & 0.422 & 31.4 \\
S2 VIs + weather & Grass & RandomForest & 1967 & mean & 0.503 & 42.7 \\
S2 VIs + weather & Shrub & RandomForest & 9565 & mean & 0.554 & 26.1 \\
S2 VIs + weather & Tree & RandomForest & 6141 & mean & 0.490 & 29.5 \\
S2 compact & Grass & XGBoost & 1967 & mean & 0.672 & 34.7 \\
S2 compact & Shrub & RandomForest & 9565 & mean & 0.675 & 22.3 \\
S2 compact & Tree & RandomForest & 6141 & mean & 0.663 & 24.0 \\
S2 compact + S1 & Grass & XGBoost & 1815 & mean & 0.661 & 35.2 \\
S2 compact + S1 & Shrub & RandomForest & 9091 & mean & 0.657 & 22.7 \\
S2 compact + S1 & Tree & RandomForest & 5608 & mean & 0.650 & 24.5 \\
L8/9 compact & Grass & XGBoost & 1186 & median & 0.691 & 35.4 \\
L8/9 compact & Shrub & RandomForest & 5440 & mean & 0.711 & 22.6 \\
L8/9 compact & Tree & RandomForest & 5030 & mean & 0.613 & 25.2 \\
L8/9 compact + thermal & Grass & XGBoost & 1186 & mean & 0.698 & 35.1 \\
L8/9 compact + thermal & Shrub & RandomForest & 5440 & mean & 0.710 & 22.6 \\
L8/9 compact + thermal & Tree & RandomForest & 5030 & mean & 0.610 & 25.3 \\
L8/9 compact + S1 & Grass & XGBoost & 845 & mean & 0.684 & 35.5 \\
L8/9 compact + S1 & Shrub & RandomForest & 3897 & mean & 0.683 & 23.3 \\
L8/9 compact + S1 & Tree & XGBoost & 3500 & mean & 0.626 & 24.6 \\
L8/9 compact + thermal + S1 & Grass & XGBoost & 845 & median & 0.691 & 35.2 \\
L8/9 compact + thermal + S1 & Shrub & RandomForest & 3897 & mean & 0.680 & 23.5 \\
L8/9 compact + thermal + S1 & Tree & XGBoost & 3500 & mean & 0.626 & 24.6 \\
VIIRS VIs & Grass & RandomForest & 3601 & mean & 0.281 & 59.1 \\
VIIRS VIs & Shrub & RandomForest & 14,710 & mean & 0.171 & 37.5 \\
VIIRS VIs & Tree & RandomForest & 13,978 & median & 0.120 & 38.4 \\
VIIRS VIs + weather & Grass & RandomForest & 3601 & mean & 0.469 & 50.8 \\
VIIRS VIs + weather & Shrub & RandomForest & 14,710 & mean & 0.534 & 28.1 \\
VIIRS VIs + weather & Tree & RandomForest & 13,978 & mean & 0.385 & 32.1 \\
VIIRS compact & Grass & XGBoost & 3601 & mean & 0.667 & 40.2 \\
VIIRS compact & Shrub & RandomForest & 14,710 & mean & 0.687 & 23.1 \\
VIIRS compact & Tree & RandomForest & 13,978 & mean & 0.599 & 25.9 \\
VIIRS compact + LAI/FPAR & Grass & XGBoost & 3164 & median & 0.671 & 39.7 \\
VIIRS compact + LAI/FPAR & Shrub & RandomForest & 13,649 & mean & 0.692 & 22.7 \\
VIIRS compact + LAI/FPAR & Tree & RandomForest & 12,696 & mean & 0.606 & 25.9 \\
\bottomrule
\end{tabular}}
\end{table}

\vspace{-20pt}

\begin{table}[H]
\caption{Paired environmental-only baseline for the central-pixel all-products Random Forest models. Rows, mean LFMC targets, folds, hyperparameters and event/bin weights are identical. Pruning and transformations are fitted separately in each training fold. Full models add common optical indices to weather, topography and cyclic day-of-year. Differences are full minus environmental-only; brackets give pointwise paired 95\% site-bootstrap intervals from 10,000 resamples. RMSE is in LFMC percentage points.}
\label{tab:environmental-baseline}
\resizebox{\textwidth}{!}{%
\begin{tabular}{lrrrrrr}
\toprule
\textbf{Vegetation} & \boldmath{$R^2_{\rm env}$} & \boldmath{$R^2_{\rm full}$} & \boldmath{$\Delta R^2$} \textbf{[95\% CI]} & \textbf{RMSE}\boldmath{$_{\rm env}$} & \textbf{RMSE}\boldmath{$_{\rm full}$} & \boldmath{$\Delta$}\textbf{RMSE [95\% CI]} \\
\midrule
Grass & 0.700 & 0.715 & 0.015 [0.004, 0.026] & 45.8 & 44.7 & $-$1.13 [$-$1.98, $-$0.30] \\
Shrub & 0.676 & 0.693 & 0.017 [0.012, 0.022] & 24.0 & 23.3 & $-$0.63 [$-$0.83, $-$0.44] \\
Tree & 0.670 & 0.691 & 0.021 [0.015, 0.029] & 26.5 & 25.6 & $-$0.86 [$-$1.18, $-$0.60] \\
\bottomrule
\end{tabular}%
}
\end{table}

\vspace{-20pt}

\begin{table}[H]
\caption{SRF compatibility and fold-safe empirical reflectance calibration for non-reference sensors in the all-products scenario. Spectral overlap is computed from normalised RTTOV/NWP SAF SRFs relative to Sentinel-2 MSI Level-2A. Production coefficients are fitted from all eligible site-suitable acquisition pairs after validation. Coefficients are fitted on footprint means; held-out RMSE uses central-pixel reflectance and pools pair instances from the outer-test diagnostics of the three promoted vegetation models; every fold-specific calibration excludes components connected to held-out field dates and acquisitions and uses reflectance only. RMSE diagnostics include the production reflectance~\mbox{clipping}.}
\label{tab:srf-overlap-calibration}

\begin{adjustwidth}{-\extralength}{0cm}
%\centering %% If there is a figure in wide page, please release command \centering, for Table, ``\textwidth" should be ``\fulllength"
\resizebox{\fulllength}{!}{%
\begin{tabular}{llrrrrrrr}
\toprule
\textbf{Sensor} & \textbf{Band} & \textbf{Production Pairs} & \textbf{Held-Out Pairs} & \boldmath{$\Delta\lambda_\mathrm{eff}$} \textbf{[nm]} & \textbf{SRF Overlap} & \textbf{Slope} & \textbf{Intercept} & \textbf{Held-Out RMSE Before/After} \\
\midrule
L8 & nir & 5280 & 2602 & 0.2 & 0.774 & 0.912 & 0.029 & 0.035/0.033 \\
L8 & red & 5280 & 2602 & $-$10.3 & 0.647 & 1.083 & $-$0.000 & 0.024/0.023 \\
L8 & swir1 & 5280 & 2602 & $-$3.3 & 0.944 & 1.020 & 0.014 & 0.036/0.031 \\
L9 & nir & 714 & 334 & 0.2 & 0.774 & 0.934 & 0.022 & 0.032/0.030 \\
L9 & red & 714 & 334 & $-$10.3 & 0.647 & 1.070 & 0.000 & 0.022/0.021 \\
L9 & swir1 & 714 & 334 & $-$3.3 & 0.944 & 1.005 & 0.016 & 0.033/0.029 \\
Terra MODIS & nir & 7058 & 3708 & $-$7.9 & 0.560 & 0.737 & 0.068 & 0.044/0.042 \\
Terra MODIS & red & 7058 & 3708 & $-$18.5 & 0.470 & 1.035 & 0.005 & 0.036/0.034 \\
Terra MODIS & swir1 & 7058 & 3708 & 17.3 & 0.468 & 0.926 & 0.025 & 0.052/0.050 \\
Aqua MODIS & nir & 2261 & 1034 & $-$7.5 & 0.562 & 0.676 & 0.079 & 0.045/0.042 \\
Aqua MODIS & red & 2261 & 1034 & $-$19.0 & 0.456 & 1.007 & 0.006 & 0.036/0.035 \\
Aqua MODIS & swir1 & 2261 & 1034 & 16.0 & 0.471 & 0.879 & 0.035 & 0.055/0.053 \\
S3A & nir & 5453 & 2730 & 1.0 & 0.920 & 0.363 & 0.129 & 0.077/0.042 \\
S3A & red & 5453 & 2730 & 0.4 & 0.364 & 0.695 & 0.019 & 0.043/0.036 \\
S3A & swir1 & 5453 & 2730 & 0.3 & 0.785 & 0.708 & 0.072 & 0.066/0.059 \\
S3B & nir & 2993 & 1526 & 1.0 & 0.920 & 0.384 & 0.131 & 0.063/0.044 \\
S3B & red & 2993 & 1526 & 0.4 & 0.364 & 0.745 & 0.019 & 0.039/0.035 \\
S3B & swir1 & 2993 & 1526 & 0.3 & 0.785 & 0.799 & 0.068 & 0.065/0.056 \\
Suomi NPP VIIRS & nir & 7056 & 3700 & $-$2.5 & 0.591 & 0.702 & 0.072 & 0.050/0.044 \\
Suomi NPP VIIRS & red & 7056 & 3700 & $-$26.8 & 0.378 & 1.018 & 0.007 & 0.041/0.041 \\
Suomi NPP VIIRS & swir1 & 7056 & 3700 & $-$10.1 & 0.758 & 0.910 & 0.034 & 0.057/0.054 \\
\bottomrule
\end{tabular}%
}
\end{adjustwidth}
\end{table}

\vspace{-20pt}

\subsection{Within-Site LFMC Retrieval}

Table~\ref{tab:unified-gapfill} reports the fixed central-pixel within-site
date-holdout configuration for each vegetation class and product scenario. In the
all-products scenario, pooled \Rtwo reaches \AllProductsGrassRtwo for Grass, \AllProductsShrubRtwo for Shrub
and \AllProductsTreeRtwo for Tree, with RMSE values of \AllProductsGrassRmse, \AllProductsShrubRmse and \mbox{\AllProductsTreeRmse percentage}
points. The corresponding no-MODIS scores are \NoModisGrassRtwo, \NoModisShrubRtwo and \NoModisTreeRtwo. These values are
not intended as single-product rankings: the no-MODIS pool has a different
site-date composition, shorter temporal depth and fewer retained observations,
particularly for Grass. Nevertheless, the MODIS record provides a measurable benefit
in extending the record and increasing the number of retained events after all products
have been mapped to the common feature space.

The controlled MODIS ablation favours inclusion of MODIS training observations
for Shrub, with an increase of 0.013 in $R^2$ and a reduction of 0.49 LFMC
percentage points in RMSE. Paired site-bootstrap intervals for Grass and Tree
include zero (Appendix~\ref{app:multisensor-diagnostics}).

In the fold-matched product-wise benchmark, multisensor training yields higher
\Rtwo and lower RMSE point estimates in all 18 product--vegetation comparisons.
Paired site-bootstrap intervals favour multisensor training for both metrics in
16 comparisons. The complete product-specific results and matched experimental
design are presented in Appendix~\ref{app:multisensor-diagnostics}.

\begin{table}[H]
\caption{Primary within-site date-holdout performance with central-pixel reflectance and mean LFMC fixed during revision. Model families are retained from the original central-pixel development comparisons and refitted with corrected acquisition-safe calibration; no new candidate selection is performed. Product rows are satellite matches and field events are unique site--date--functional-class targets. Metrics give each field event equal total weight. The no-MODIS scenario excludes both MODIS products.}
\label{tab:unified-gapfill}
\resizebox{\textwidth}{!}{%
\begin{tabular}{lllrrrrr}
\toprule
\textbf{Scenario} & \textbf{Vegetation} & \textbf{Model} & \textbf{Product Rows} & \textbf{Field Events} & \boldmath{$R^2$} & \textbf{RMSE} & \textbf{KGE} \\
\midrule
All Products & Grass & RandomForest & 13,506 & 5741 & 0.715 & 44.7 & 0.779 \\
All Products & Shrub & RandomForest & 67,437 & 25,340 & 0.693 & 23.3 & 0.756 \\
All Products & Tree & RandomForest & 37,976 & 15,062 & 0.691 & 25.6 & 0.801 \\
No MODIS & Grass & XGBoost & 6182 & 3128 & 0.645 & 43.5 & 0.793 \\
No MODIS & Shrub & RandomForest & 37,207 & 17,584 & 0.680 & 24.2 & 0.753 \\
No MODIS & Tree & RandomForest & 17,368 & 8148 & 0.689 & 25.3 & 0.799 \\
\bottomrule
\end{tabular}%
}
\end{table}

\vspace{-3pt}

\begin{figure}[H]
\includegraphics[width=0.99\textwidth]{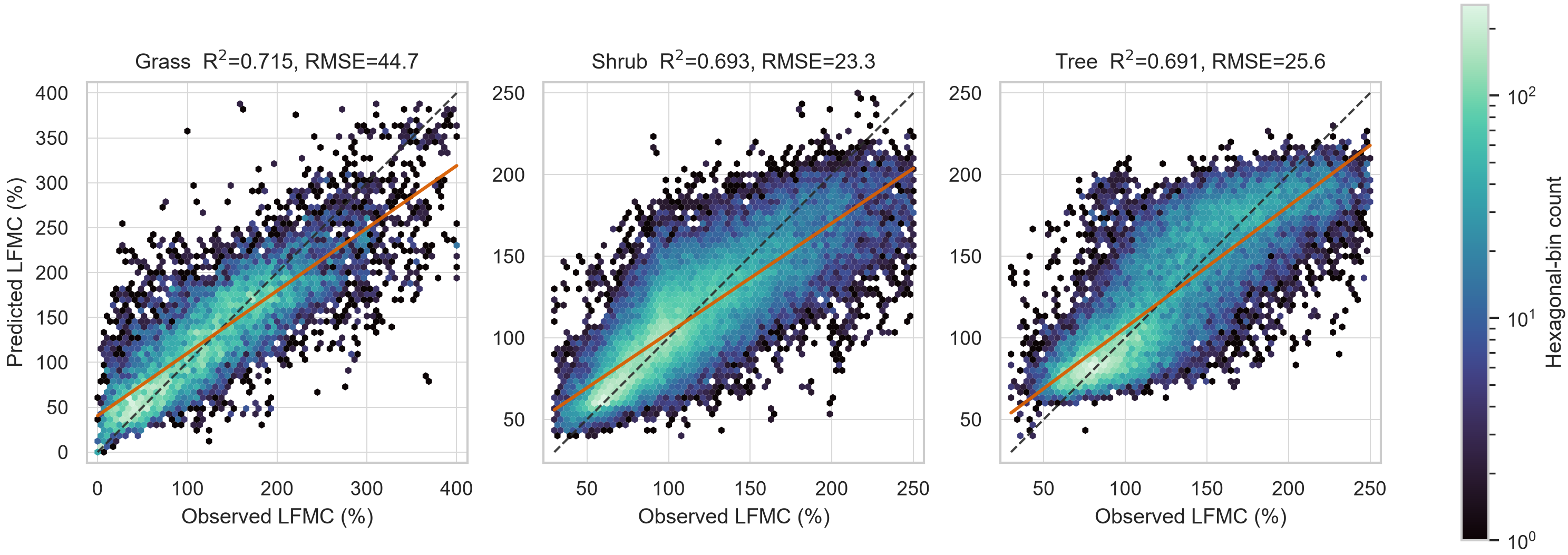}
\caption{Observed versus predicted LFMC for the final central-pixel all-products primary within-site date-holdout model in each vegetation class. Points are plotted as hexagonal density bins, with brighter colours indicating denser
observed-predicted regions. The diagonal line marks perfect agreement. Metrics
in the panel titles are pooled across held-out folds. The orange line is the
ordinary least-squares fit between observed and predicted LFMC and is included
to show the empirical response slope of the model predictions.}
\label{fig:unified-gapfill-scatter}
\end{figure}

Figure~\ref{fig:unified-gapfill-scatter} shows observed versus predicted LFMC
for the final central-pixel all-products within-site models. Grass covers the widest
LFMC range, while Shrub and Tree predictions are more concentrated because their target domains are narrower
and their observed distributions are strongly peaked. The scatter plots also
show a compressed prediction range for woody vegetation: very low LFMC
observations tend to be overestimated, while very high LFMC observations tend
to be underestimated. This indicates that the models capture broad LFMC
gradients but dampen the observed amplitude of site-level moisture variation,
consistent with the limited and indirect sensitivity of top-of-canopy optical
reflectance to live fuel water status \citep{yebra2013review,quan2021}.
In fire-danger applications, overestimating low LFMC could understate the
contribution of dry live fuels to hazard, whereas underestimating high LFMC
could overstate it. The effect on warning thresholds should therefore be
evaluated within the intended fire-danger system before operational use.

\vspace{-3pt}

All final models use central-pixel reflectance and the fixed event-level
mean LFMC outcome. In the matched all-products sensitivity, neighbourhood
means and medians increase $R^2$ by 0.0028--0.0048 and reduce RMSE by
0.13--0.26 LFMC percentage points relative to central pixels. Paired intervals
favour both summaries for Shrub and Tree and include zero for Grass.
Central-pixel predictors were retained to simplify inference on each
product-native grid; the modest neighbourhood gains and complete comparisons
are reported in Supplementary File S5.

Under the tested WorldCover profiles, absolute changes from the production
profile are at most 0.016 in $R^2$ and 0.49 LFMC percentage points in RMSE.
Removing the LFMC variability cutoff substantially expands the woody-vegetation
cohorts and reduces Tree $R^2$ from 0.691 to 0.534. These comparisons describe
sensitivity to the retained population because eligibility changes also alter
evaluation support and folds (Appendix~\ref{app:filter-sensitivity};
\mbox{Supplementary File S5}).

\subsection{Predictor-Importance Diagnostics}

The final models, obtained by pooling all the products and trained and validated
using the within-site, date holdout validation mode, are Random Forest models for all three
vegetation classes. This allows a compact diagnostic based on impurity-based
feature importances. These importances are not interpreted as causal effects:
correlated predictors can split importance, and tree-based importances favour
variables that create useful recursive partitions. They are nevertheless
useful for interpreting how much predictive weight is carried by the main
predictor families: spectral vegetation indices, weather, topography and
seasonality.

\begin{figure}[H]
\includegraphics[width=0.99\textwidth]{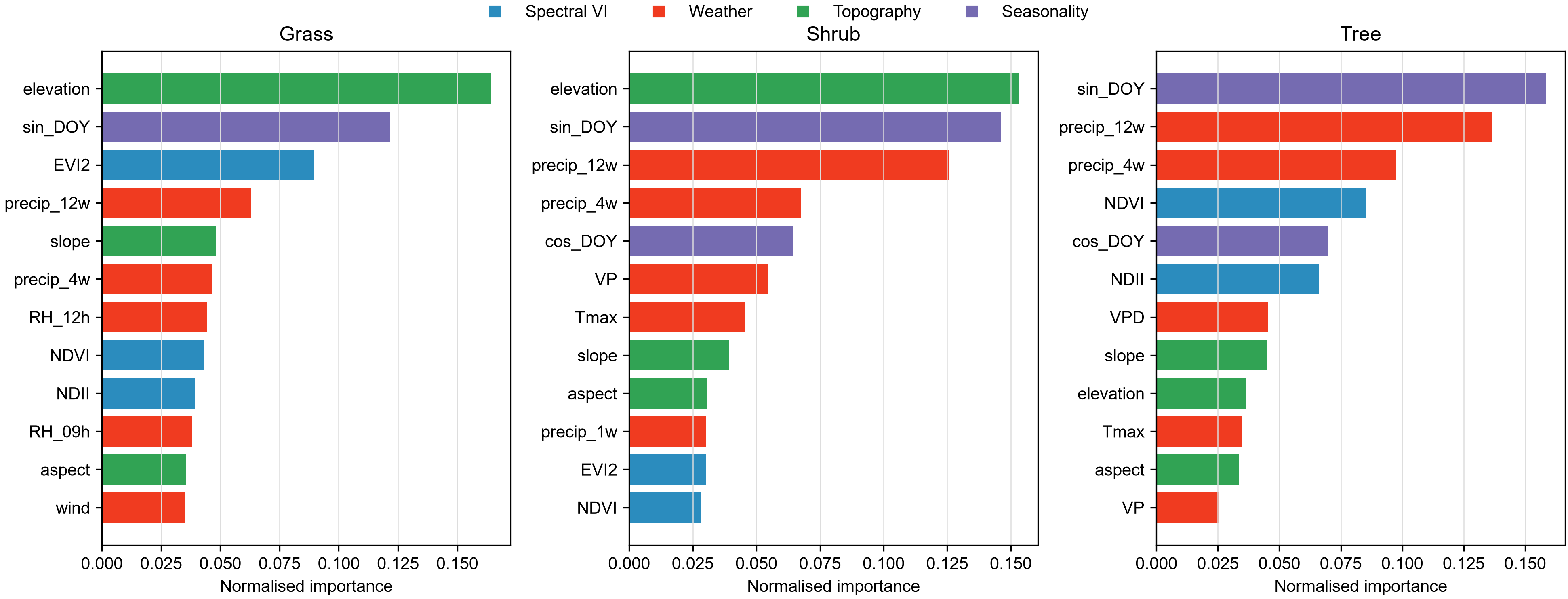}
\caption{Top predictors in the final all-products primary within-site
date-holdout models.
Bars show normalised Random Forest feature importances for the exported best
Grass, Shrub and Tree models. Colours identify predictor families. The figure
is used as a diagnostic of model behaviour, not as a causal attribution of LFMC
controls.}
\label{fig:unified-feature-importance}
\end{figure}

Table~\ref{tab:unified-feature-importance-groups} and
Figure~\ref{fig:unified-feature-importance} show class-specific use of the
central-pixel compact feature set. For Grass, grouped importances are
weather (0.428), topography (0.248), spectral vegetation indices (0.172), seasonality (0.152). For Shrub, grouped importances are
weather (0.484), topography (0.223), seasonality (0.210), spectral vegetation indices (0.083). For Tree, grouped importances are
weather (0.505), seasonality (0.228), spectral vegetation indices (0.151), topography (0.115). These diagnostics
describe the fitted models; the paired environmental-only comparison
quantifies the incremental predictive contribution of optical information.

\begin{table}[H]
\caption{Grouped feature importances for the final all-products primary within-site date-holdout models. Importances are impurity-based Random Forest importances normalised within each vegetation class and grouped by predictor family. Values are diagnostic rather than causal, because correlated predictors can share importance.}
\label{tab:unified-feature-importance-groups}
\begin{tabularx}{\textwidth}{@{\extracolsep{\fill}}llr@{}}
\toprule
\textbf{Vegetation} & \textbf{Feature Group} & \textbf{Normalised Importance} \\
\midrule
Grass & Weather & 0.428 \\
Grass & Topography & 0.248 \\
Grass & Spectral VI & 0.172 \\
Grass & Seasonality & 0.152 \\
Shrub & Weather & 0.484 \\
Shrub & Topography & 0.223 \\
Shrub & Seasonality & 0.210 \\
Shrub & Spectral VI & 0.083 \\
Tree & Weather & 0.505 \\
Tree & Seasonality & 0.228 \\
Tree & Spectral VI & 0.151 \\
Tree & Topography & 0.115 \\
\bottomrule
\end{tabularx}
\end{table}

The low grouped spectral vegetation-index importance for Shrub should not be interpreted as evidence that optical indices convey little information about shrub LFMC. The single-product screening indicates that vegetation indices alone retain useful information, especially for Sentinel-2 MSI Level-2A and Landsat 8/9 Collection 2 Level-2. In the final all-products Random Forest model, Shrub retrieval is instead more strongly influenced by weather, topography and seasonality. This likely reflects the discontinuous structure of many shrublands, where optical footprints contain mixed canopy, soil/background and terrain signals, especially for moderate-resolution products. In addition, impurity-based importances can redistribute importance among correlated predictors and predictor families. The result can therefore be interpreted as a reduced contribution of the common optical indices after environmental context is included, rather than as low useful VI information content, and this is also shown by the single-product experiments carried out during the preliminary feature-set screening.

\vspace{+2pt}

\section{Discussion}
\label{sec:discussion}

\subsection{Role of Environmental Context in LFMC Retrieval}

The final framework can be interpreted as a machine-learning LFMC retrieval
model driven by satellite vegetation indices, weather, topography and
seasonality. The satellite variables provide information on greenness and
canopy water-sensitive reflectance. The meteorological variables provide
antecedent rainfall and atmospheric-demand context. Topography helps represent
site-scale environmental gradients, and day of year captures phenological
seasonality. The common optical feature space is therefore most informative
when interpreted together with meteorological, topographic and seasonal
context. The results show that a compact spectral feature set, when combined
with environmental predictors and trained on a large field archive, can
estimate LFMC with useful accuracy across
functional vegetation classes. The feature-screening experiment
(Table~\ref{tab:feature-ablation-gapfill}) makes this point explicit: spectral
vegetation indices alone provide better performance for higher-resolution sensors, whereas adding weather, topography
and seasonality produces most of the final retrieval skill.

The paired environmental baseline (Table~\ref{tab:environmental-baseline})
shows that optical indices add predictive information after weather, topography
and seasonality are included. The size of this increment depends on the training
regime (Appendix~\ref{app:optical-pooling-ablation}). Product-specific models
show larger optical $R^2$ gains in 16 of the 18 sensor--vegetation combinations,
with pointwise paired intervals supporting nine between-regime differences.
The largest Grass increments occur with Landsat and Sentinel-2. However, pooled
full models retain higher $R^2$ on identical held-out observations in all 18
comparisons. Pooled training strengthens the environmental baseline and overall
retrieval performance, while reducing the relative optical increment in several
settings. This comparison retains the same fold-specific reflectance calibration
and accounts for changes in training support and weighting; it does not isolate
harmonisation as the cause. The broader field-matched archive and extended
temporal coverage directly serve the framework's objective of making fuller
use of GlobeLFMC observations across products and resolutions.

\subsection{Screening of Optional Product-Specific Predictors}

Several potentially useful variables were tested before fixing the final
multisensor feature space. Sentinel-1 radar backscatter could in principle add
structural and moisture information under clouds \citep{torres2012sentinel1,
rao2020sar}; Landsat thermal information could capture canopy and
surface temperature stress \citep{masek2020landsat9}; VIIRS LAI/FPAR could
constrain canopy cover in a way similar to previous MODIS-based approaches
\citep{nasa_vnp15a2h_002, quan2021}.
In our screened branches, however, these variables did not provide enough
accuracy gain to compensate for the loss of sample support and the additional
product-specific dependencies. In particular, adding predictors from another
satellite data product can require valid observations from both products within
the prescribed temporal matching windows. A field record with a usable optical
observation can then be excluded if no quality-screened observation is available
for the additional product within its matching window. The final compact set is therefore not a
minimalist assumption made in advance: it is the result of an explicit
screening step showing that most LFMC predictability in these experiments is
captured by common vegetation indices, weather, topography and seasonality.
This greatly simplifies sensor unification because the same predictor
definitions can be computed for MODIS, VIIRS, Landsat, Sentinel-2 MSI Level-2A
and Sentinel-3 SY\_2\_SYN 300 m.
A separate optical-free Sentinel-1 diagnostic is presented in
Appendix~\ref{app:sentinel1-optical-free}; it evaluates a complementary
predictor set that operates without optical observations.

\subsection{Effect of Multisensor Integration on Model Performance}

The unified design links fuller use of GlobeLFMC 2.0 with a common LFMC
estimation tool for products operating at different spatial resolutions.
Combining product archives makes more field observations available for
training and validation and reduces the dependence of model development on a
single sensor's observation period. MODIS adds observations from the early
2000s and increases Grass and Shrub support, while VIIRS, Landsat,
Sentinel-2 MSI Level-2A and Sentinel-3 SY\_2\_SYN 300 m provide more recent
continuity and complementary spatial support. The resulting architecture is
intended to let researchers and fire managers select a product according to
the spatial detail and temporal coverage required by their application.
The controlled comparison in Table~\ref{tab:controlled-modis-ablation}
provides evidence of a modest MODIS training benefit for Shrub, with low impact
on Grass and Tree. However, MODIS also provides the benefit of expanding the
historical archive and temporal coverage of field-matched observations,
independently of its incremental predictive effect on recent non-MODIS records.
The fold-matched sensor-wise comparison in
Table~\ref{tab:sensorwise-matched-benchmark} provides complementary evidence:
relative to models trained on one product only, the unified training set
yields favourable point estimates in every product--vegetation combination.
The paired 95\% intervals favour multisensor training for both metrics in
\mbox{16 comparisons}. For Tree with Sentinel-2 and Sentinel-3, the intervals for the
performance differences include zero, which corresponds to equal performance
under product-specific and multisensor training. Although multisensor training
improves the point estimates in these two comparisons, the gains are not clearly
distinguished from the variation estimated by resampling whole sites.

The footprint sensitivity in Supplementary File S5 shows modest gains from
neighbourhood averaging on identical valid observations. The final model uses
central-pixel reflectance to provide a simpler product-native implementation.
The surrounding 3-by-3 land-cover and valid-pixel checks remain part of site
suitability: they assess whether each central observation lies within a
credible vegetated environment at the product's native scale. Larger supports
for moderate-resolution products are screened using the corresponding
product-specific rules. Field-coordinate uncertainty and the representativeness
of the reported point remain sources of uncertainty.

The same workflow can be extended to additional optical sensors when the
required information is available. A new product must provide red, NIR and
SWIR1 reflectance observations, documented SRFs, product-specific spatial
support information, and enough temporally overlapping observations with the
existing archive to fit the target-independent reflectance calibration and
evaluate the resulting model. Under those conditions, the new product can be
sampled with its own spatial footprint, calibrated toward the
Sentinel-2 reference feature domain, filtered for site suitability and evaluated
by comparing model performance with and without that product in the training
archive. The paper's contribution is therefore not limited to the LFMC model
and to a subset of GlobeLFMC 2.0 expanded with satellite observations, 
but also includes a reusable multisensor workflow that can be exploited
to further expand satellite sensor support.

\subsection{Sensor-Specific Uses of the Unified Framework}

The unified framework supports both retrospective analyses and future
monitoring through complementary sensor options. A pre-fire mapping example
for the 2018 Camp Fire in California compares Sentinel-2 and Sentinel-3 applications on
their 20 m and nominal 300 m grids (Appendix~\ref{app:camp-fire-case}).

MODIS serves two complementary purposes: it expands the field-matched
training archive and enables retrospective LFMC reconstruction over more
than two decades, beginning in 2000. It provides the longest record among
the products supported here. This historical coverage has scientific value
independently of future data acquisition, supporting analyses of
interannual variability, drought episodes and past fire seasons. MODIS is
therefore retained primarily for historical and legacy applications.
The National Aeronautics and Space Administration (NASA) projects the end of science activities for Terra in February 2027 and
Aqua in September 2027 \citep{nasa2025terraend}. These anticipated retirements
limit future observations but do not diminish the utility of the archived
record. Future monitoring can use the supported VIIRS and Sentinel-3
products, alongside Landsat and Sentinel-2 where finer spatial detail is
required.

Sentinel-2 and Landsat 8/9 offer the best spatial resolution 
and a relatively long observation archive, making them ideal for local applications, wildland--urban
interface studies and regions where fine spatial detail is more important than large spatial coverage
or \mbox{revisit frequency}.

VIIRS VNP09H1 provides a convenient 8-Day composite
at a 500 m resolution which is ideal for large-scale (i.e., regional, national and continental) 
applications and which can easily be integrated into Google Earth Engine-based workflows.
VIIRS surface-reflectance products were developed to provide continuity
with MODIS \citep{nasa2026viirscontinuity}.

Sentinel-3 Synergy provides a 300 m moderate-resolution branch that is better aligned with near-real-time
continental monitoring that benefits from or requires daily updates.

The September 2026 launch of
Sentinel-3C, the third satellite in the Sentinel-3 series, also supports
future mission continuity following commissioning \citep{esa2026s3claunch};
its observations were not included in the present study.
Supporting multiple products provides a route to continued LFMC monitoring,
but the current validation does not establish a seamless long-term LFMC
series across sensor transitions. Such continuity requires assessment of
residual sensor-dependent differences, changes in spatial support and
product latency for the intended application.

\subsection{Comparison with Previous Work}

Previous MODIS-based LFMC studies demonstrated that global or regional LFMC
mapping is feasible from moderate-resolution optical data
\citep{cunill2022rf, quan2021}. Quan et~al.~\citep{quan2021} reported global
MODIS LFMC retrieval with an all-class \Rtwo of 0.62 without additional site
filtering and 0.71 under their optimal temporal/homogeneity filter. Their
class-wise optimal-filter results were \Rtwo = 0.75 for grassland, 0.59 for
shrubland and 0.45 for forest, evaluated on 765, 408 and 835 validation
measurements, respectively. The present all-products within-site models obtain
\Rtwo = \AllProductsGrassRtwo, \AllProductsShrubRtwo and \AllProductsTreeRtwo for Grass, Shrub and Tree, respectively. Evaluation uses \AllProductsGrassEvents, \AllProductsShrubEvents and \AllProductsTreeEvents unique held-out field events, represented by \AllProductsGrassProductRows, \AllProductsShrubProductRows and \AllProductsTreeProductRows satellite-product rows. The comparison is not a controlled
replication because the field archive, predictor set, filtering rules,
products and validation design differ, but it shows that the unified empirical
branch is competitive with previous global MODIS work while using a much
larger GlobeLFMC 2.0-derived validation set.

Cunill Camprub\'i et~al.~\citep{cunill2022rf} produced a Mediterranean-basin
LFMC product using Random Forests with MODIS spectral, thermal and seasonal
predictors. Their model used 10,374 field samples from GlobeLFMC and the
Catalan LFMC monitoring programme, with reported RMSE values of 19.9\% for
calibration samples from 2000 to 2014 and 16.4\% for validation samples from
2015 to 2019. That regional product is not directly comparable to a global
multisensor GlobeLFMC 2.0 model, but it supports the broader conclusion that
tree-based empirical methods can retrieve LFMC effectively when representative
field data and environmental predictors are available.

Wang et~al.~\citep{wang2026tcn} combined 365-day MODIS and Parameter-elevation Regressions on Independent Slopes
Model (PRISM) spectral--meteorological sequences with terrain variables
in a temporal convolutional network (TCN) coupled with a bidirectional
gated recurrent unit (BiGRU) network for daily LFMC estimation over the contiguous United States. Their
model-ready archive contained 66,469 MODIS-pixel-day samples from 931 sites
after averaging coincident field observations within each pixel and date. In
their principal temporal experiment, 21,524 observations from later years
formed the independent evaluation set; the model obtained a Pearson
correlation coefficient of $R=0.75$ and RMSE of 25.02\%. This is a relevant
recent comparison because it also combines optical and meteorological
information over a long field archive, but its reported $R$ is not
interchangeable with the coefficient of determination reported here. Its
contiguous-US, single-sensor, 365-day sequence design also addresses a
different retrieval setting from the global, multisensor, event-based models
developed in the present study.

Cho et~al.~\citep{cho2025smap} evaluated a complementary passive-microwave
approach at five National Fuel Moisture Database sites in California and
Oregon. Vegetation water content and soil moisture were retrieved from Soil Moisture Active Passive (SMAP)
L-band brightness temperature using a Maximum Entropy Production formulation,
with 300~m fractional vegetation cover used to represent sub-grid vegetation
within the 9~km SMAP grid. Correlations between field LFMC and retrieved
vegetation water content at the five sites were $r=0.81$, 0.62, 0.80, 0.05 and
0.90. The study therefore demonstrates the potential of microwave-derived
vegetation water content as an LFMC proxy, but it does not report a pooled
\Rtwo, RMSE or independent LFMC regression validation and should not be
interpreted as a direct performance benchmark for the present models.

Rao et~al.~\citep{rao2020sar} used Sentinel-1 backscatter and Landsat-8
reflectance in a physics-assisted recurrent neural network to map LFMC every
15 days at 250 m over the western United States.~Their cross-validation at 125
sites produced \Rtwo = 0.63, RMSE = 25.0\% and \mbox{bias = 1.9\%}, and adding
microwave backscatter improved \Rtwo from 0.44 to 0.63.~This illustrates the
potential value of radar information for regional high-resolution LFMC
mapping. In the present multisensor experiments, however, Sentinel-1 features
did not improve the representative high-resolution branch sufficiently to
justify a radar dependency in the final global common-feature model. One
plausible explanation is that antecedent precipitation and atmospheric-demand
predictors already encode much of the moisture-history signal that C-band
backscatter could otherwise contribute at the site scale. This interpretation
remains a hypothesis because the experiment does not isolate substitution of
radar for weather predictors. A separate optical-free branch combines
Sentinel-1 backscatter and incidence angle with weather, topography and cyclic
day of year (Appendix~\ref{app:sentinel1-optical-free}). Its fixed central-pixel
Random Forest models achieve $R^2=0.571$, 0.628 and 0.670 for Grass, Shrub and
Tree, respectively, on a separate Sentinel-1-matched cohort. The paired
comparison favours the environmental-only baseline for Grass and Shrub; the
Tree difference remains uncertain. The branch therefore demonstrates retrieval
without an optical acquisition requirement, while these tests do not establish
an accuracy gain from radar. The optical-free branch may nevertheless be
useful where frequent or persistent cloud cover prevents regular, complete
optical acquisitions, including parts of northern Europe and Canada, because
Sentinel-1 observations are available through cloud cover
\citep{torres2012sentinel1}. A further research direction is to test physics-informed
Sentinel-1 models, motivated by the approach of Rao \mbox{et al.\ \citep{rao2020sar}},
and evaluate their LFMC retrieval skill and incremental contribution relative
to the environmental-only baseline.

Yebra et~al.~\citep{yebra2026highres} demonstrated Sentinel-2 MSI Level-2A
LFMC monitoring across Australia with strong grassland performance and lower
forest and shrubland skill. In their Australian Flammability Monitoring System (AFMS) emulation test, grassland agreement
was high (\Rtwo = 0.83, RMSE = 32.45\%), whereas forests and shrublands showed
lower skill (\Rtwo = 0.43, RMSE = 20.84\%; and \Rtwo = 0.21, RMSE = 10.28\%,
respectively). Their independent GlobeLFMC 2.0 validation improved at
homogeneous sites (\Rtwo = 0.42, \mbox{RMSE = 31.39\%}) and reached \Rtwo = 0.53
with a dedicated Sentinel-2 validation campaign. Their study also emphasised
the importance of site homogeneity for remote-sensing validation. The present
work generalises that idea into product-specific site suitability and LFMC
time-series adequacy filters, then extends the modelling framework beyond a
single high-resolution product to MODIS, VIIRS, Landsat 8/9, Sentinel-2 and Sentinel-3 Synergy.
Yebra \mbox{et al.\ \citep{yebra2026highres}} also found broadly consistent
validation performance across field--satellite matching windows of
$\pm3$, $\pm5$ and $\pm10$ days. This finding provides context for the
bounded matching windows used here, although it does not establish timing
insensitivity for every product or vegetation class in the present study.

\subsection{Limitations and Future Work}

Several limitations remain. First, GlobeLFMC 2.0 is the largest available
global field LFMC archive, but observations are still concentrated in specific
countries, ecosystems and monitoring programmes. This affects both model
training and validation, especially for under-represented biomes. Pooled
Grass, Shrub and Tree metrics summarise performance over the retained
observations and do not establish equal skill across biomes or species
within a class. Product-specific site-suitability filtering addresses
land-cover mixing, but does not ensure balanced ecological representation;
biome- and species-specific validation remains a priority for future work.
Complementary site-blocked validation yields lower skill in all three classes
than temporal validation on the same records
(Appendix~\ref{app:site-blocked-validation}). This provides evidence for
prediction at additional sites within the environments represented in the
archive.

The geographical distribution of the field measurements also limits the
accuracy assessment of continental or global applications. Any model validated
against GlobeLFMC alone remains subject to the regions and vegetation conditions
represented in that archive. Combining satellite records with different
observation periods and spatial resolutions makes fuller use of the available
field data and facilitates application over large areas wherever the required
inputs are available. However, the ability to produce an LFMC map does not
establish its accuracy throughout the mapped area. For regions or environmental
conditions that are not adequately represented by the field observations,
performance cannot be quantified from the present validation; independent
field measurements are needed to assess accuracy there.

Second, 
optical reflectance is primarily sensitive to the upper canopy and illuminated 
background within the satellite footprint, whereas field LFMC may describe leaves, 
shoots or species components that are not equally visible from above. 
This mismatch can only be partly addressed by the site-suitability filter.

Sentinel-3 SYNERGY already incorporates cloud and snow screening during
product generation \citep{copernicus2026s3synprocessing}. The training sampler
uses the resulting finite reflectances without further screening of the
supplied quality flags. Additional screening could exclude cloud margins,
partly cloudy pixels and unreliable retrievals that retain finite values,
and remains a possible quality-control improvement. The Camp Fire mapping
example applies additional cloud and snow flag screening
(Appendix~\ref{app:camp-s2-s3}).

The site-suitability filter itself is also an operational approximation because GlobeLFMC does not
provide field polygons. WorldCover can identify obviously mixed or
non-vegetated footprints, but it cannot prove that the exact field plot
occupies the central satellite pixels. The static WorldCover layer also cannot
capture all land-cover changes that occurred throughout the 2000--2023
satellite archive. Future work could assess site suitability separately for each
site and observation date using temporally matched Landsat imagery and, where
available, Sentinel-2 imagery to characterise vegetation composition within the
sampled footprints. Such time-varying screening could improve the reliability
of the filter, particularly for coarse-resolution sensors, but would require
additional image retrieval, quality screening and vegetation classification
across the archive, substantially increasing the processing and computational
workload.

The terrain-adjusted weather used in the Camp Fire example
(Appendix~\ref{app:camp-fire-case}) uses a reconstructed historical HRES
elevation field. Its exact equivalence to AgERA5's original reference grid is
unconfirmed, leaving uncertainty in the local temperature and humidity
adjustments. The adjustment is applied during inference with the final trained LFMC
models; the example provides no independent assessment of its effect on accuracy. The training
archive and reported validation metrics retain the original AgERA5 predictors.

The harmonisation is limited to robust linear overlap calibration. More flexible calibration approaches could be tested, but they
must remain target-independent to avoid learning LFMC from the calibration
pairs. Finally, the strict common feature space excludes bands available only
to some products, such as blue, green, red-edge or the second shortwave-infrared band (SWIR2). Future work should
test whether a hierarchical framework can retain the common core while adding
sensor-specific optional features where they improve predictions without
breaking transferability.

Future work should also evaluate sensors and sampling strategies that address
limitations of optical top-of-canopy reflectance. Radar observations from the
NASA--Indian Space Research Organisation Synthetic Aperture Radar (NISAR)
mission and the BIOMASS mission may be particularly useful in woody
vegetation. These missions are so recent that GlobeLFMC observations are not yet
available to test retrieval approaches that incorporate their measurements, 
but they are relevant for future operational extensions. Sentinel-1 operates at C band, whereas NISAR
provides L- and S-band observations and BIOMASS provides P-band observations;
these longer wavelengths can penetrate deeper into woody canopies and may carry information
on structure and moisture conditions that optical reflectance does not capture
\citep{nasa2026nisar, esa2026biomass}.

For fire-danger applications, the
final product should be evaluated as an input to fire-danger rating systems,
including the next generation of the Daily Fire Danger Index~\citep{pampanoni2023thesis}, to test whether direct satellite LFMC estimates
improve fire-danger skill relative to meteorological drought proxies alone.

\section{Conclusions}

We present a unified multisensor machine-learning framework designed to
make fuller use of GlobeLFMC 2.0 for training and validation and to provide
the remote-sensing and fire-management communities with a common LFMC
estimation tool operating at different spatial resolutions. It integrates
Terra and Aqua MODIS, VIIRS, Landsat 8/9, Sentinel-2 and Sentinel-3 observations
through shared optical and environmental predictors, product-specific site
suitability and target-independent reflectance calibration.

Under primary within-site date-holdout validation, the all-products model reaches pooled
\Rtwo values of \AllProductsGrassRtwo for Grass, \AllProductsShrubRtwo for Shrub and \AllProductsTreeRtwo for Tree.

Optical indices provide additional skill above the environmental baseline,
although their incremental contribution is smaller under pooled training in
several sensor--vegetation combinations. Product-specific fits show larger
optical increments in some settings, while the pooled full models retain higher
$R^2$ point estimates on identical test observations in all 18 comparisons.
The common architecture therefore combines complementary optical information
with the predictive benefit and broader coverage of a larger training archive.

The resulting workflow provides, to our knowledge, the first common
multisensor LFMC architecture designed to retain the spatial support of each
input product. Satellite observations provide direct information on the current
state of vegetation that complements weather, topography and seasonality.
They also allow LFMC estimates to reflect spatial differences in vegetation at
the resolution of the selected satellite product, including the 20--30~m
resolution of Sentinel-2 and Landsat, which is substantially finer than the
weather inputs. Together with the 300--500~m products, these native grids
address the spatial requirements of local fuel monitoring and broader
fire-danger analyses.

The primary scores apply to additional dates at represented sites;
complementary site-blocked validation, as expected, demonstrates lower skill at unseen sites
within the sampled domains (Appendix~\ref{app:site-blocked-validation}).
The models can be applied over large areas wherever the required inputs are
available. However, their accuracy in regions or vegetation conditions insufficiently
represented in GlobeLFMC remains unquantified by the present evaluation unless new
observations become available.
We consider this multisensor approach well suited to making the fullest
possible use of the currently available GlobeLFMC observations, because
combining satellite records extends both the observation period and the range
of environmental conditions represented in training. Independent field
measurements remain necessary to quantify performance in regions and
vegetation conditions insufficiently represented by GlobeLFMC.

This sensor-agnostic architecture
requires harmonisation, spatial support and site suitability to remain explicit
parts of the methodology. Additional optical sensors can be added when they
provide red, NIR and SWIR1 reflectance bands, documented SRFs and enough
overlapping observations to support calibration and validation.

Following publication, the authors will publicly release the final trained
models, user documentation, a companion Jupyter notebook and scripts for
using the models to support LFMC research and fire-management applications.

\vspace{6pt}
\supplementary{ {The} %MDPI: 1. We added necessary link in this section according to our requirements. Please confirm. 2. The descriptions of the supplementary files in this section are not totally the same with the provided file names. Please check if they should be revised to keep consistent. @VP: Confirmed the supplementary-link format. Filenames and descriptions now match all seven files in the ZIP downloaded from the MDPI submission portal.
 following supporting information can be downloaded at \linksupplementary{s1}: Supplementary File S1: site inclusion and exclusion decisions (\nolinkurl{Supplementary_S1_site_inclusion_and_exclusion.xlsx});
Supplementary File S2: per-site model diagnostics (\nolinkurl{Supplementary_S2_per_site_model_diagnostics.xlsx});
\mbox{Supplementary File S3:} field--satellite matching and satellite-sampling protocol (\nolinkurl{Supplementary_S3_satellite_sampling_protocol.pdf});
Supplementary File S4: site-bootstrap uncertainty, paired model comparisons and supporting diagnostics (\nolinkurl{Supplementary_S4_site_bootstrap_uncertainty.xlsx});
Supplementary \mbox{File S5}: LFMC-target, satellite-footprint and filter sensitivities (\nolinkurl{Supplementary_S5_target_and_footprint_sensitivity.xlsx});
Supplementary File S6: reflectance-harmonisation and LFMC sensitivity diagnostics (\nolinkurl{Supplementary_S6_harmonization_sensitivity.xlsx});
and Supplementary File S7: optical-free Sentinel-1 comparison and predictor-importance diagnostics (\nolinkurl{Supplementary_S7_optical_free_Sentinel1.xlsx}).}

\authorcontributions{Conceptualisation, V.P., E.C. and G.L.; methodology, V.P., A.F. and S.S.; software, V.P.; validation, V.P. and A.F.; formal analysis, V.P. and A.F.; data curation, V.P. and A.F.; visualisation, V.P.; supervision, E.C. and G.L.; writing---original draft preparation, V.P.; w\mbox{riting---review} and editing, V.P., E.C., A.F., G.L. and S.S. All authors have read and agreed to the published version of the manuscript.}

\funding{ {This research was carried out} %MDPI: Information regarding the funder and the funding number should be provided. Please check the accuracy of funding data and any other information carefully. Any updates after publication should be carefully considered. @VP: checked, and added "No." between "CUP" and the project number consistently with our recent email exchange.
 within the Space It Up! project, funded by the Italian Space Agency and the Italian Ministry of University and Research under contract 2024-5-E.0, CUP No. I53D24000060005.}

\institutionalreview{Not applicable.}

\informedconsent{Not applicable.}

\dataavailability{The GlobeLFMC 2.0 field database and the satellite, meteorological and topographic datasets analysed in this study are available from the providers cited in the manuscript. The derived analysis tables, complete run configurations, and trained Grass, Shrub and Tree models are available to editors and reviewers from the corresponding author upon reasonable request. Following publication, the authors will publicly release the final trained models, user documentation, a companion Jupyter notebook and scripts for using the models. Additional workflow components may be released in response to community interest. The Supplementary Materials provide site-inclusion tables, per-site model diagnostics and the satellite-sampling protocol.}

\acknowledgments{The authors acknowledge the GlobeLFMC contributors and the providers of the NASA, ESA, Copernicus and Google Earth Engine datasets used in this study. During preparation of this manuscript, the authors used OpenAI Codex to assist with code review and reproducibility checks. All scientific decisions, analyses and final text were reviewed and approved by the authors.}

\conflictsofinterest{ {Alvise Ferrari was employed by GMATICS S.r.l. and Simone Saquella was employed by Serco Italia S.p.A. The remaining authors declare that the research was conducted in the absence of any commercial or financial relationships that could be construed as a potential conflict \mbox{of interest.}}} %MDPI: There are authors from company (affs. 3 and 4), it is required that they disclose the relevant potential interest relations (refer to https://www.mdpi.com/ethics#_bookmark17). Please revise the statement to the following contents: @VP: Revised the statement to disclose employment of Alvise Ferrari at GMATICS S.r.l. and Simone Saquella at Serco Italia S.p.A., as requested. Their contributions to this study were made pro bono, without compensation for those contributions.
%"Author XXXXXXX was employed by the company XXXXXXX. The remaining authors declare that the research was conducted in the absence of any commercial or financial relationships that could be construed as a potential conflict of interest".

\abbreviations{Abbreviations}{%
~The following abbreviations are used in this manuscript:\\

{\noindent
\begin{tabular}{p{2cm}p{8cm}}
CDSE & Copernicus Data Space Ecosystem\\
DOY & Day of year\\
FPAR & Fraction of absorbed photosynthetically active radiation\\
%\end{tabular}}
%
%{\noindent
%\begin{tabular}{p{2cm}p{8cm}}
KGE & Kling--Gupta efficiency\\
LAI & Leaf area index\\
LFMC & Live fuel moisture content\\
MODIS & Moderate Resolution Imaging Spectroradiometer\\
NIR & Near infrared\\
RMSE & Root mean square error\\
SRF & Spectral response function\\
SWIR & Shortwave infrared\\
VIIRS & Visible Infrared Imaging Radiometer Suite
\end{tabular}}}

\appendixtitles{yes}
\appendixstart
\appendix
\section{Complementary Site-Blocked Validation}
\label{app:site-blocked-validation}

This diagnostic evaluates prediction at GlobeLFMC sites absent from LFMC model
training and reflectance calibration. It complements the primary within-site
date-holdout analysis while retaining the final all-products eligible records,
arithmetic-mean LFMC targets, central-pixel predictors and fixed Random Forest
hyperparameters. Eligibility is defined by the existing full-series suitability
and adequacy criteria. Historical model-family development choices are retained;
this is a conditional diagnostic and does not constitute an independent
post-selection evaluation.

Five site folds are generated once, with a fixed random seed of 42, using
stratified grouped allocation on one row per unique field event. Vegetation
class is the stratification variable and the GlobeLFMC site identifier is the
grouping variable; LFMC values and model performance are not used to choose
folds. A single assignment is shared by all three classes, so every eligible
record from a site belongs to the same fold. No geographic buffer, regional
holdout or environmental-domain blocking is imposed.

For each fold, every observation from the held-out sites is excluded from the
reflectance calibration pool, including records outside the class-specific
LFMC modelling cohorts. Remaining quality-screened, site-suitable reflectance
pairs are used to refit the mean-derived calibration. Eligible reflectance-only
sites outside the LFMC modelling cohorts can contribute calibration pairs.
The same calibration is applied to all three class models in that fold.
Feature pruning, scaling, event-balanced LFMC-bin weights and target
transformation are estimated using each class model's training observations
only. Every retained product row receives exactly one held-out prediction.
Field-event weights sum to one across the products associated with that event.
Production models and the primary temporal validation \mbox{remain unchanged}.

Table~\ref{tab:site-blocked-validation} compares site-held-out predictions with
the primary temporal predictions on identical product rows, targets and event
weights. Pointwise 95\% intervals use 10,000 whole-site bootstrap resamples
with seed 42, conditional on the fixed cohorts, folds, configurations and fitted
predictions; models are not refitted within bootstrap draws. Full uncertainty,
site assignments and fold support are supplied in Supplementary File S4.

\begin{table}[H]
\caption{Complementary site-blocked validation and primary within-site temporal validation on identical all-products records. Support is given as sites/unique field events/product rows. Brackets indicate site-bootstrap 95\% intervals. RMSE is in LFMC percentage points. All models use central-pixel predictors, mean LFMC targets and fixed Random Forest settings.}
\label{tab:site-blocked-validation}

\begin{adjustwidth}{-\extralength}{0cm}
%\centering %% If there is a figure in wide page, please release command \centering, for Table, ``\textwidth" should be ``\fulllength"
%\resizebox{!}{
\begin{tabularx}{\fulllength}{LlRrRr}
\toprule
\textbf{Vegetation} & \textbf{Sites/Events/Rows} & \boldmath{$R^2_{\rm temporal}$} & \boldmath{$R^2_{\rm site}$} \textbf{[95\% CI]} & \textbf{RMSE}\boldmath{$_{\rm temporal}$} & \textbf{RMSE\boldmath{$_{\rm site}$} [95\% CI]} \\
\midrule
Grass & 89/5741/13,506 & 0.715 & 0.529 [0.430, 0.615] & 44.7 & 57.4 [51.9, 63.4] \\
Shrub & 274/25,340/67,437 & 0.693 & 0.567 [0.516, 0.608] & 23.3 & 27.7 [26.1, 29.5] \\
Tree & 125/15,062/37,976 & 0.691 & 0.596 [0.533, 0.649] & 25.6 & 29.3 [27.6, 31.1] \\
\bottomrule
\end{tabularx}
%}
\end{adjustwidth}
\end{table}

The five folds hold out the following numbers of sites per class: Grass: 15--19; Shrub: 54--55; Tree: 22--26.

Holding out sites reduces $R^2$ by 0.186, 0.126 and 0.095 for Grass, Shrub
and Tree, respectively, and increases RMSE by 12.73, 4.38 and 3.66 LFMC
percentage points. Paired intervals support lower skill under site holdout in
all three classes. Positive site-held-out $R^2$ provides evidence of transfer
within the archive's sampled domains, with a substantial loss of accuracy when
site context is absent from training. Nearby sites can share climatic and
ecological conditions; holding out site identifiers does not establish transfer
to unseen regions, biomes or environmental domains. Geographic bias and
conditional record selection remain relevant to both validation modes.

\section{Camp Fire Pre-Fire LFMC Mapping}
\label{app:camp-fire-case}

The 2018 Camp Fire in Butte County, California, provides a retrospective
application example. CAL FIRE records its start on 8 November at 06:33
Pacific Standard Time (14:33 UTC) \citep{calfire2018camp}.
The reference perimeter is obtained from the publicly hosted Camp Fire
perimeter layer \citep{camp2019perimeter}. It supplies a geographic outline
and a region for summaries and is not an LFMC predictor. The mapped extent
also retains surrounding vegetation.

\begin{figure}[H]
\includegraphics[width=\textwidth]{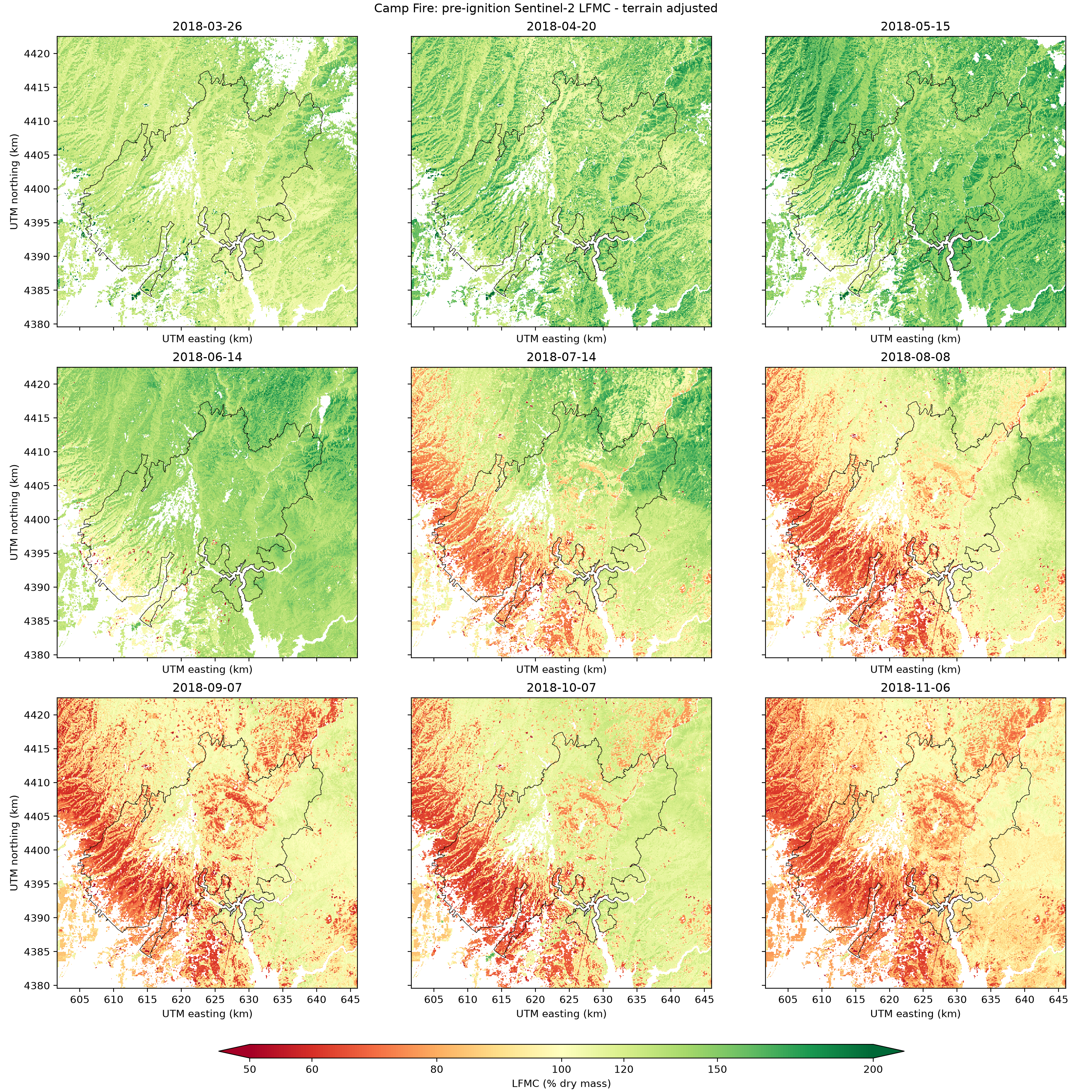}
\caption{Estimated pre-fire LFMC for the Camp Fire region on nine Sentinel-2 acquisition dates in 2018, using terrain-adjusted AgERA5 and the final all-products models. Panels share the native \mbox{20 m} optical grid, extent and colour scale. The black outline marks the eventual fire perimeter; white areas have unsupported dominant land cover, masked optical observations or incomplete predictors. Vegetation routing uses a fixed Dynamic World probability composite preceding the first acquisition. Red denotes LFMC at or below 50\%, yellow 100\% and dark green at or above 200\%; these display limits do not clip prediction values and do not define universal fire-danger thresholds. All acquisitions precede ignition on 8 November.}
\label{fig:camp-lfmc-maps}
\end{figure}

Nine Sentinel-2 Level-2A acquisition dates from 26 March to 6 November 2018
illustrate seasonal variation before ignition
(Figure~\ref{fig:camp-lfmc-maps}). Dates provide approximately monthly coverage,
subject to cloud-screened availability; the March acquisition was selected
for the lowest mean tile cloud fraction among March scenes covering the region.
Red reflectance is area-averaged from 10 m onto the native 20 m B8A/B11 grid.
Central-pixel indices, ALOS terrain, daily AgERA5 weather and seasonal predictors
follow the final model definitions. Cloud, cirrus, shadow, snow and invalid
scene-classification categories are masked, and predictions require complete
central-pixel features. The final all-products Random Forest models predict
mean LFMC without additional fitting or local field-data calibration.

\subsection{Pre-Fire Vegetation Routing and Terrain-Adjusted Weather}

WorldCover 2020 postdates this event. Vegetation routing therefore uses
Dynamic World, a Sentinel-2-derived 10 m land-cover probability product
\citep{brown2022dynamicworld}. Probabilities for all nine classes are averaged
over 26 March 2017--25 March 2018, preceding the earliest mapped acquisition,
and area-averaged onto the 20 m optical grid. The dominant class selects the
Grass, Shrub or Tree model for grass, shrub-and-scrub or trees, respectively;
other dominant classes remain unmapped. This fixed pre-sequence composite
avoids incorporating post-fire observations or changing routing between dates.
Training-site purity, neighbourhood homogeneity and neighbouring-valid-pixel
thresholds are not applied as inference masks.

All maps and time-series summaries use terrain-adjusted AgERA5 weather
with the final trained LFMC models. We first bilinearly interpolate daily
weather and reference elevation to the optical pixel centres, then adjust
weather for the difference between local ALOS elevation and reference height.
Interpolation requires valid source values at every neighbouring node with a
positive weight; missing neighbours are not replaced or assigned zero weight.
AgERA5 is
bias-corrected toward the European Centre for Medium-Range Weather Forecasts
(ECMWF) high-resolution operational model (HRES); its product guide describes
an additional local-elevation temperature correction using a 6.5 K/km lapse
rate~\citep{ecmwf2026agera5guide}.~To reconstruct reference elevation, archived
HRES surface geopotential for 1~April 2016 is obtained from the NSF National
Center for Atmospheric Research (NCAR) Geoscience Data Exchange
\citep{ecmwf2016hresarchive}. The archived regular-longitude/Gaussian-latitude
field is bilinearly interpolated using its actual coordinates to the AgERA5
0.1-degree nodes and divided by 9.80665 m/s$^2$ to obtain geopotential height.
Representative fields from model cycles 41r2, 43r1 and 43r3 have identical
terrain values. Exact equivalence to the original AgERA5 provider elevation
grid remains unconfirmed because its original terrain-interpolation and
masking procedure could not be established from the available documentation.
This is a known limitation of the terrain-adjusted inference example: the
reconstructed reference height introduces an unquantified contribution to the
local temperature and humidity adjustments. Geopotential height is used as
an approximation to geometric height when comparing the reference elevation
with the local ALOS DEM.

Elevation corrections combining a temperature lapse rate with unchanged
relative humidity have precedent in the North American Land Data Assimilation
System and ERA5-Land \citep{cosgrove2003nldas, munoz2021era5land}.
The fixed 6.5 K/km lapse rate is a commonly used approximation;
actual lapse rates vary with location and atmospheric conditions
\citep{dutra2020lapserate}.

Let $\Delta z$ denote local ALOS elevation minus interpolated reference
height, and \linebreak  \mbox{$\Gamma=0.0065$ K/m}. Daily mean and maximum temperature are adjusted
as $T'=T-\Gamma\Delta z$. With $e_s$ denoting the saturation vapour pressure
function used in the training workflow, define
$q=e_s(T'_{\mathrm{mean}})/e_s(T_{\mathrm{mean}})$.
Daily vapour pressure is adjusted as $\mathrm{VP}'=q\,\mathrm{VP}$, preserving
its daily relative-humidity proxy. Daily dewpoint is adjusted separately as
$T'_d=e_s^{-1}[q\,e_s(T_d)]$, since mean dewpoint and mean vapour pressure are
distinct daily aggregates. VPD is recomputed from adjusted daily mean
temperature and vapour pressure. Rainfall, wind and relative humidity at
individual hours retain their interpolated source values. Missing inputs or
thermodynamic combinations inconsistent with this approximation are masked.

We apply this approach to daily aggregates, preserving the ratios
$\mathrm{VP}/e_s(T_{\mathrm{mean}})$ and
$e_s(T_d)/e_s(T_{\mathrm{mean}})$ as separate humidity proxies.
Because saturation vapour pressure is nonlinear in temperature, these ratios
do not generally equal daily mean relative humidity, and $e_s(T_d)$ calculated
from mean dewpoint does not generally equal daily mean vapour pressure. The daily adjustment
therefore approximates an hourly correction followed by daily aggregation.

The adjustment is applied once before inference, using the local optical
grid or native geolocated Sentinel-3 centres. It does not modify the training
archive or constitute an independently validated improvement in LFMC accuracy.

\begin{figure}[H]
\includegraphics[width=\textwidth]{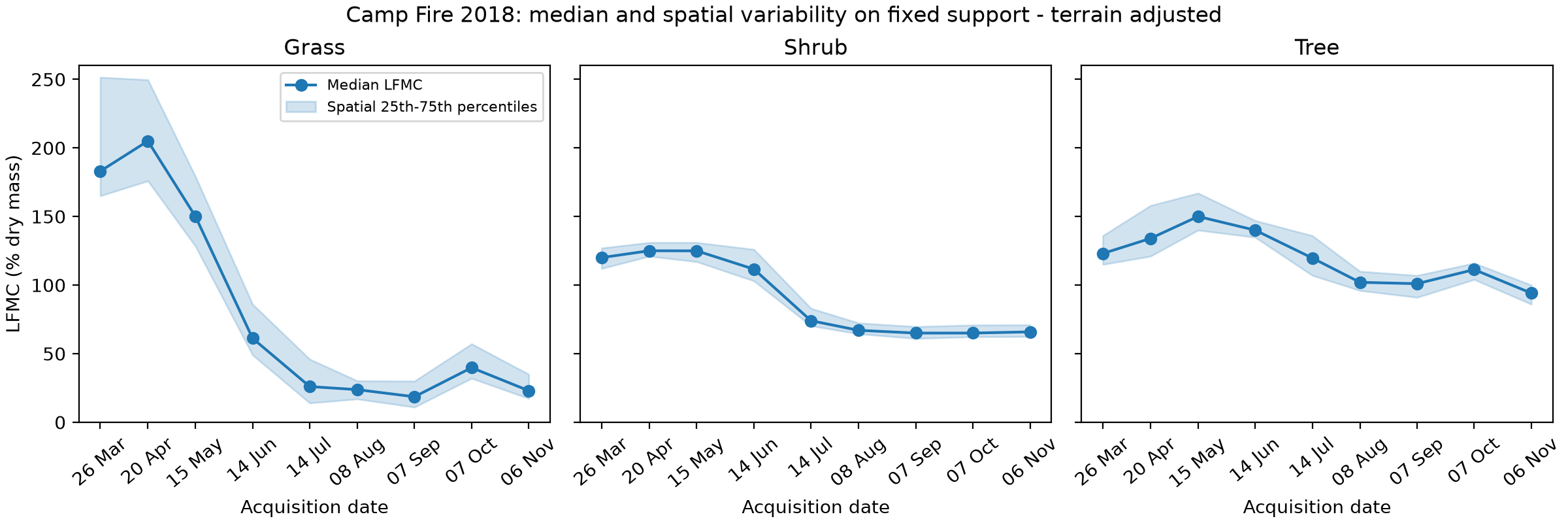}
\caption{Vegetation-class median LFMC and spatial interquartile ranges from terrain-adjusted inference inside the eventual Camp Fire perimeter. The common support retains identical pixels with valid terrain-adjusted predictions across all nine acquisitions. Predictions use fixed pre-sequence Dynamic World routing and frozen models. Shading represents spatial 25th--75th percentiles, not prediction confidence intervals. Lines connect acquisition dates and do not represent daily LFMC~\mbox{estimates}.}
\label{fig:camp-lfmc-timeseries}
\end{figure}

Figure~\ref{fig:camp-lfmc-timeseries} summarises terrain-adjusted LFMC on identical
pixels inside the eventual perimeter, valid across all nine acquisition dates. This support comprises 348.12 ha of Grass, 13,631.80 ha of Shrub and
39,501.76 ha of Tree. The longer sequence reveals substantial seasonal drying:
terrain-adjusted median LFMC decreases from 183.0\%, 120.0\% and 123.0\% on
26 March to 23.0\%, 65.9\% and 94.1\% on 6 November for Grass, Shrub and Tree,
respectively. The trajectories are not monotonic: spring maxima and an October
recovery are evident. These summaries describe spatial LFMC variation within
the mapped vegetation classes; local field measurements would be needed to
assess their accuracy.

\subsection{Native-Resolution Sentinel-2 and Sentinel-3 Comparison}
\label{app:camp-s2-s3}

\begin{figure}[H]
\includegraphics[width=\textwidth]{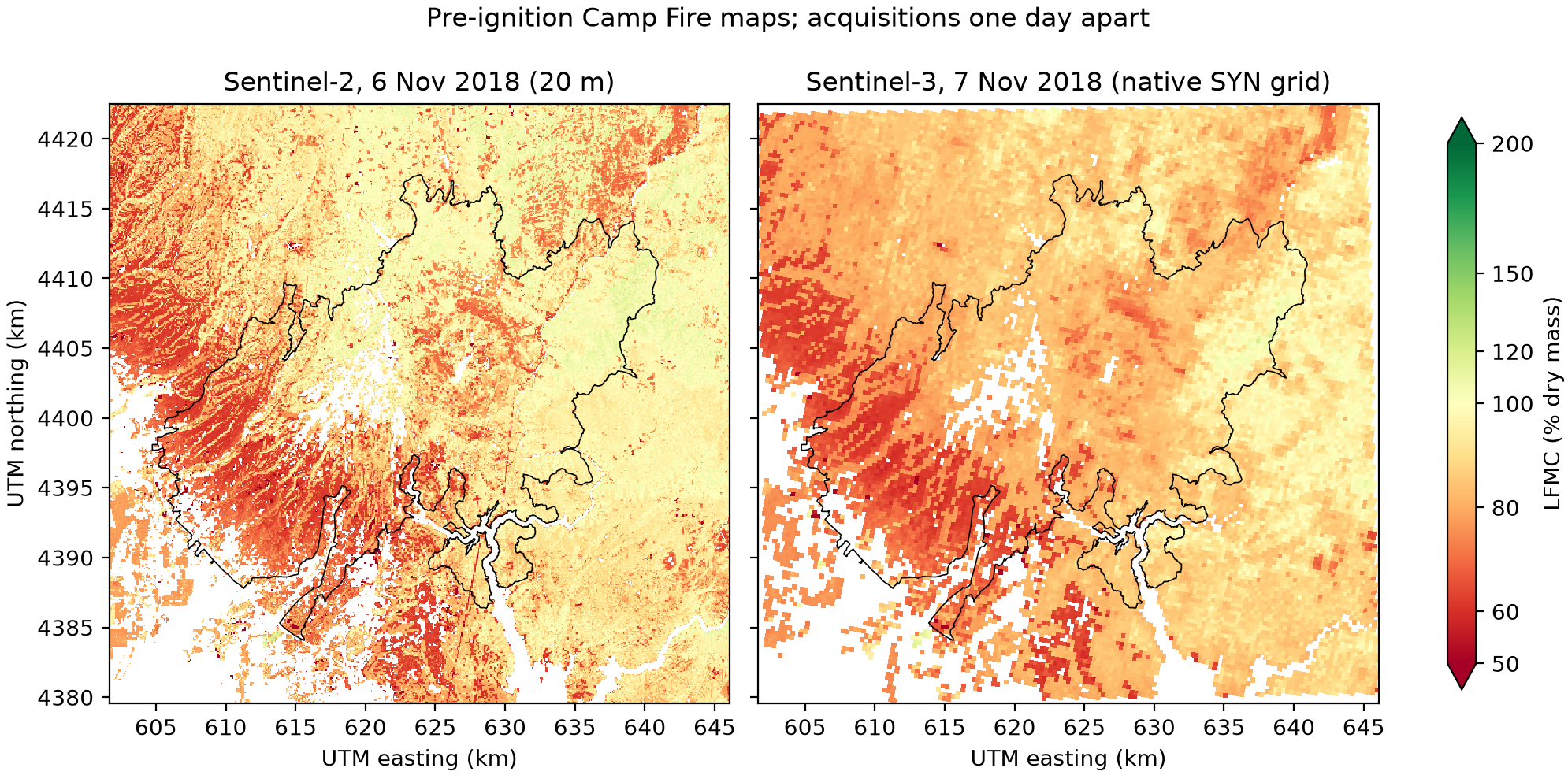}
\caption{Pre-fire LFMC in the Camp Fire region from Sentinel-2 on 6 November 2018 (20 m) and Sentinel-3A SY\_2\_SYN on 7 November (original geolocated nominal 300 m grid). Both panels share the geographic extent, LFMC scale, fixed pre-sequence vegetation composite, terrain-adjustment method and final all-products class models. Daily weather and optical acquisitions correspond to each date. The eventual fire perimeter is outlined. Sentinel-3 includes collocated SLSTR SWIR information; no LFMC smoothing or downscaling is applied. White areas indicate unmapped pixels.}
\label{fig:camp-s2-s3}
\end{figure}

A Sentinel-3A SY\_2\_SYN acquisition on 7 November 2018 supplies a
moderate-resolution example, one day after the latest Sentinel-2 acquisition
and before ignition (Figure~\ref{fig:camp-s2-s3}). Red, NIR and SWIR1 reflectance
are taken from OLCI Oa08, OLCI Oa17 and collocated SLSTR S5N, respectively, on the
original geolocated nominal 300 m OLCI grid~\citep{copernicus2023s3syn}.
The final Sentinel-3A-specific calibration precedes central-pixel index
calculation. No reflectance or LFMC downscaling is applied. Dynamic World
probabilities from the same fixed composite are area-averaged over nominal
300 m central support before selecting the vegetation model. Central cloud,
cloud-margin, ambiguous-cloud and snow flags are masked, and complete predictors
are required. Weather and reference elevation are bilinearly sampled at the
native centres; the same terrain-adjustment method and frozen all-products
models are used. The non-time-critical product was processed after ignition,
so this remains a retrospective application.

Within the perimeter, Sentinel-3 supplies 11 Grass, 1485 Shrub and 5329 Tree
predictions, with terrain-adjusted medians of 42.7\%, 66.0\% and 84.0\%,
respectively. Both products display broad spatial contrasts, while Sentinel-2
resolves finer vegetation patches. Differences combine spatial support,
spectral response, viewing geometry, routing and the one-day acquisition
offset. In particular, the small Grass sample limits class-level comparison.
The maps do not isolate resolution as an independent experimental factor.

The case study demonstrates retrospective spatial application at two native
optical resolutions and seasonal variation using daily weather analyses.
Corresponding local field LFMC measurements are unavailable; transfer accuracy
and the terrain adjustment therefore require independent validation in the
intended domain. 

\section{Controlled Multisensor Training Diagnostics}
\label{app:multisensor-diagnostics}

\subsection{Contribution of MODIS Training Observations}

The independent all-products and no-MODIS scenarios describe performance on
their respective eligible records. To isolate the contribution of MODIS training
observations, a controlled ablation retains the all-products eligible cohort,
mean LFMC targets, central-pixel predictors, Random Forest hyperparameters and
original acquisition-blocked folds. Within each fold, Terra and Aqua MODIS
rows are removed from one training arm. Both models are evaluated on identical
non-MODIS held-out rows at sites represented in both training arms. Five product
rows from four site/fold combinations are excluded from both evaluation arms
because those sites have no non-MODIS training observations in the corresponding
fold. Event weights are recomputed within each training arm and the shared
non-MODIS evaluation subset. Pruning, scaling, LFMC-bin weighting and target
transformation follow the same procedures estimated separately from each
training arm. Both arms use the same acquisition-safe, target-independent
fold-local reflectance calibration. This evaluates the contribution of MODIS
training observations conditional on a common harmonised feature space;
calibration itself is not ablated. Paired uncertainty is estimated with 10,000
whole-site bootstrap resamples. Full support and exclusion details are in
Supplementary File S4.

The controlled comparison in Table~\ref{tab:controlled-modis-ablation}
separates MODIS training contribution from differences in evaluation support.
Adding MODIS increases $R^2$ by 0.004 for Grass and 0.013 for Shrub and changes
Tree $R^2$ by $-$0.002, with corresponding RMSE changes of $-$0.28, $-$0.49
and +0.09 LFMC percentage points. Paired site-bootstrap intervals favour
MODIS inclusion for Shrub; the Grass and Tree intervals include zero.

\subsection{Matched Product-Specific and Multisensor Training}

A second, fold-matched diagnostic evaluates whether multisensor training
improves retrieval within each individual sensor domain. For every product and
vegetation class, a product-specific Random Forest is fitted using only that
product's rows from each outer training fold. It is compared with the promoted
all-products Random Forest on the identical held-out rows from that product.
The two fits use identical target definitions, model hyperparameters and
fold-specific reflectance calibration. Calibration uses the full eligible
pairing pool after excluding field dates and acquisition-connected components
associated with the held-out observations. Correlation pruning, predictor
scaling, LFMC-bin weighting and quantile target transformation follow the same
procedures but are estimated separately within each model's training subset.
The diagnostic therefore evaluates the effect of pooling training products
conditional on a common multisensor calibration.

\begin{table}[H]
\caption{Controlled MODIS training ablation on identical non-MODIS held-out observations at sites represented in both training arms. Mean LFMC targets, central-pixel predictors, Random Forest hyperparameters, acquisition-blocked folds and fold-local reflectance calibration are shared. Pruning, scaling, event/bin weighting and target transformation follow identical procedures fitted separately within each training arm. Differences are all-products minus no-MODIS; brackets give paired 95\% site-bootstrap intervals from 10,000 resamples. RMSE is in LFMC percentage points.}
\label{tab:controlled-modis-ablation}

\begin{adjustwidth}{-\extralength}{0cm}
%\centering %% If there is a figure in wide page, please release command \centering, for Table, ``\textwidth" should be ``\fulllength"
%\resizebox{!}{
\begin{tabularx}{\fulllength}{lrRrrrr}
\toprule
\textbf{Vegetation} & \boldmath{$R^2_{\rm no\text{-}MODIS}$} & \boldmath{$R^2_{\rm all}$} & \boldmath{$\Delta R^2$} \textbf{[95\% CI]} & \textbf{RMSE}\boldmath{$_{\rm no\text{-}MODIS}$} & \textbf{RMSE}\boldmath{$_{\rm all}$} & \boldmath{$\Delta$}\textbf{RMSE [95\% CI]} \\
\midrule
Grass & 0.661 & 0.665 & 0.004 [$-$0.007, 0.016] & 45.1 & 44.8 & $-$0.28 [$-$1.07, 0.48] \\
Shrub & 0.673 & 0.685 & 0.013 [0.008, 0.018] & 24.4 & 24.0 & $-$0.49 [$-$0.66, $-$0.32] \\
Tree & 0.681 & 0.678 & $-$0.002 [$-$0.010, 0.004] & 25.5 & 25.6 & 0.09 [$-$0.17, 0.37] \\
\bottomrule
\end{tabularx}
%}
\end{adjustwidth}
\end{table}

The product-wise evaluation in Table~\ref{tab:sensorwise-matched-benchmark}
uses the final all-products model but reports its held-out predictions separately
for each sensor. To provide a controlled baseline, a second model is trained for
each product and vegetation class using only that product's training rows. Both
models are evaluated on exactly the same product-specific test rows and the same
acquisition-blocked folds; consequently, differences cannot be attributed to a
different test set. With central-pixel predictors, multisensor training yields higher \Rtwo and
lower RMSE point estimates in all 18 comparisons. Paired site-bootstrap
intervals favour multisensor training for both metrics in 16
comparisons. The Grass \Rtwo gains range from 0.017 to
0.102. The Shrub \Rtwo gains range from 0.019 to
0.065. The Tree \Rtwo gains range from 0.014 to
0.049. Comparisons whose intervals include zero remain uncertain
under site resampling (Supplementary File S4).

\subsection{Optical Contribution Under Pooled and Product-Specific Training}
\label{app:optical-pooling-ablation}

To quantify how training support affects the incremental optical contribution,
we extend the matched product-wise comparison with an environmental-only
baseline for each product and vegetation class. The baseline includes weather,
topography and cyclic day of year; the full model additionally includes the
common optical indices. Each product-specific pair uses only that product's
rows from the original outer training folds. Its predictions are compared with
the pooled environmental and full models on identical held-out product rows,
mean LFMC targets and acquisition-blocked fold assignments. The original
fold-specific reflectance calibration and Random Forest hyperparameters are
retained. Predictor pruning, scaling, event-balanced LFMC-bin weights and target
transformation are estimated separately within each training subset. Field-event
weights are recalculated within the evaluated product so each field event has
unit total weight. Thus, this diagnostic evaluates the effect of product pooling
and its associated training support and weighting, conditional on the existing
reflectance calibration. Sentinel-2 is the calibration reference and receives
no fitted cross-sensor reflectance adjustment.

Table~\ref{tab:optical-pooling-ablation} reports the optical increment for both
training regimes. Differences between their increments are evaluated using
10,000 paired whole-site bootstrap resamples of the fixed predictions, with
seed 42 and pointwise 95\% intervals. Models are not refitted within bootstrap
draws. Complete baseline and full-model metrics, intervals for optical increments
and their between-regime differences, per-site diagnostics and fold support are
provided in Supplementary File S4.

\begin{table}[H]
\caption{Fold-matched central-pixel single-product versus multisensor Random Forest comparison on identical held-out rows, mean LFMC targets and acquisition-blocked folds. Fold calibration is shared; predictive-model preprocessing is fitted separately. Entries list single/multisensor metrics. Positive $\Delta R^2$ and negative $\Delta$RMSE favour multisensor training. Brackets give pointwise paired 95\% site-bootstrap intervals from 10,000 resamples. RMSE is in LFMC percentage points. \mbox{Supplementary File S4} gives complete intervals and support.}
\label{tab:sensorwise-matched-benchmark}

\begin{adjustwidth}{-\extralength}{0cm}
%\centering %% If there is a figure in wide page, please release command \centering, for Table, ``\textwidth" should be ``\fulllength"
\resizebox{\fulllength}{!}{%
\begin{tabular}{llrcccc}
\toprule
\textbf{Product} & \textbf{Vegetation} & \textbf{Sites} & \boldmath{$R^2$} \textbf{Single/Multi} & \boldmath{$\Delta R^2$} \textbf{[95\% CI]} & \textbf{RMSE Single/Multi} & \boldmath{$\Delta$}\textbf{RMSE [95\% CI]} \\
\midrule
Landsat 8/9 C2 L2 & Grass & 58 & 0.555/0.657 & 0.102 [0.076, 0.136] & 47.6/41.8 & $-$5.82 [$-$7.44, $-$4.37] \\
Landsat 8/9 C2 L2 & Shrub & 210 & 0.644/0.680 & 0.036 [0.023, 0.051] & 24.6/23.3 & $-$1.27 [$-$1.74, $-$0.84] \\
Landsat 8/9 C2 L2 & Tree & 97 & 0.648/0.672 & 0.024 [0.009, 0.040] & 25.9/25.0 & $-$0.89 [$-$1.48, $-$0.33] \\
Terra MODIS & Grass & 80 & 0.706/0.723 & 0.017 [0.008, 0.028] & 46.0/44.6 & $-$1.36 [$-$2.08, $-$0.68] \\
Terra MODIS & Shrub & 228 & 0.672/0.690 & 0.019 [0.012, 0.026] & 24.0/23.3 & $-$0.70 [$-$0.92, $-$0.48] \\
Terra MODIS & Tree & 110 & 0.676/0.694 & 0.018 [0.013, 0.024] & 26.2/25.5 & $-$0.74 [$-$0.94, $-$0.56] \\
Aqua MODIS & Grass & 80 & 0.613/0.699 & 0.086 [0.054, 0.125] & 51.1/45.0 & $-$6.05 [$-$8.10, $-$4.09] \\
Aqua MODIS & Shrub & 227 & 0.625/0.690 & 0.065 [0.053, 0.080] & 25.3/23.0 & $-$2.31 [$-$2.68, $-$1.97] \\
Aqua MODIS & Tree & 110 & 0.633/0.682 & 0.049 [0.037, 0.063] & 27.5/25.6 & $-$1.89 [$-$2.35, $-$1.44] \\
Sentinel-2 MSI L2A & Grass & 46 & 0.620/0.666 & 0.046 [0.020, 0.078] & 42.5/39.9 & $-$2.67 [$-$4.26, $-$1.14] \\
Sentinel-2 MSI L2A & Shrub & 213 & 0.639/0.664 & 0.025 [0.011, 0.040] & 24.9/24.0 & $-$0.88 [$-$1.35, $-$0.40] \\
Sentinel-2 MSI L2A & Tree & 98 & 0.664/0.679 & 0.015 [$-$0.004, 0.035] & 25.2/24.6 & $-$0.55 [$-$1.29, 0.13] \\
Sentinel-3 SYN 300 m & Grass & 37 & 0.523/0.625 & 0.102 [0.047, 0.162] & 46.5/41.2 & $-$5.28 [$-$8.30, $-$2.42] \\
Sentinel-3 SYN 300 m & Shrub & 198 & 0.607/0.665 & 0.058 [0.041, 0.077] & 26.1/24.1 & $-$2.00 [$-$2.57, $-$1.46] \\
Sentinel-3 SYN 300 m & Tree & 90 & 0.629/0.653 & 0.024 [$-$0.002, 0.054] & 27.0/26.1 & $-$0.88 [$-$1.96, 0.08] \\
VIIRS VNP09H1 & Grass & 58 & 0.645/0.670 & 0.025 [0.004, 0.046] & 46.8/45.1 & $-$1.69 [$-$2.95, $-$0.29] \\
VIIRS VNP09H1 & Shrub & 220 & 0.642/0.676 & 0.034 [0.024, 0.045] & 25.5/24.3 & $-$1.23 [$-$1.56, $-$0.91] \\
VIIRS VNP09H1 & Tree & 104 & 0.667/0.682 & 0.014 [0.005, 0.024] & 26.0/25.5 & $-$0.57 [$-$0.93, $-$0.19] \\
\bottomrule
\end{tabular}%
}
\end{adjustwidth}
\end{table}

\vspace{-6pt}

Product-specific training yields a larger optical $R^2$ increment in
16 of the 18 product--vegetation combinations; intervals support a larger
increment in nine. For Grass, the largest increments occur with Landsat and
Sentinel-2: 0.074 and 0.055 under product-specific training, compared with
0.005 and 0.011 under pooling. Their RMSE reductions are 3.80 and 2.96 LFMC
percentage points under product-specific training, compared with 0.31 and 0.64
under pooling. The corresponding between-regime differences in optical
$R^2$ gain are 0.069 [0.020, 0.125] and 0.044 [0.011, 0.087].
For Shrub, intervals support larger product-specific optical increments for
Aqua MODIS, Landsat, Sentinel-2 and Sentinel-3. For Tree, they support larger
increments for Terra and Aqua MODIS; the Landsat and Sentinel-2 differences
remain uncertain. The pattern varies among vegetation classes and products.

Pooled full-model $R^2$ remains higher in all 18 comparisons, consistent with
the matched benchmark above.~The larger relative role of optical predictors
under some product-specific fits coexists with a stronger environmental
baseline and better overall point-estimate performance under pooling.
These comparisons account for changes in training support and weighting and
do not isolate reflectance harmonisation as their cause. Products also differ
in observed sites and dates, so differences among products do not establish a
\mbox{resolution effect}.

Restricting training to one product leaves 23 of the 118,919 held-out product
rows at sites without training observations from that product, although those
sites are represented in pooled training. Repeating the matched comparisons
after excluding these rows preserves which between-regime optical-gain
intervals exclude zero. This sensitivity and affected site/fold assignments
are included in Supplementary File S4.

\begin{table}[H]
\caption{ {Incremental} %MDPI: We removed the space between rows but add borders instead. Please confirm. @VP: this looks better indeed, thanks.
 effect of optical indices above weather, topography and cyclic day of year under pooled (P) and single-product (S) training. Entries list P/S on identical held-out rows for each product and vegetation class. Positive values indicate an increase in $R^2$ or a reduction in RMSE and MAPE. RMSE reductions are in LFMC percentage points; MAPE reductions are in percentage points of relative error. Complete baseline and full-model metrics, paired 95\% intervals and support are provided in Supplementary File S4.}
\label{tab:optical-pooling-ablation}
\resizebox{\textwidth}{!}{
\begin{tabular}{llrrr}
\toprule
\textbf{Product} & \textbf{Vegetation} & \boldmath{$\Delta R^2$} \textbf{P/S} & \textbf{RMSE Reduction P/S} & \textbf{MAPE Reduction P/S} \\
\midrule
Terra MODIS & Grass & 0.017/0.016 & 1.32/1.20 & 3.98/5.48 \\
Terra MODIS & Shrub & 0.020/0.024 & 0.74/0.88 & 0.43/0.71 \\
Terra MODIS & Tree & 0.020/0.028 & 0.82/1.13 & 0.69/0.95 \\
\midrule
Aqua MODIS & Grass & 0.016/0.037 & 1.17/2.38 & 3.04/7.45 \\
Aqua MODIS & Shrub & 0.017/0.026 & 0.61/0.85 & 0.37/0.71 \\
Aqua MODIS & Tree & 0.017/0.028 & 0.68/1.01 & 0.60/0.80 \\
\midrule
VIIRS & Grass & 0.018/0.020 & 1.19/1.31 & 1.99/1.91 \\
VIIRS & Shrub & 0.018/0.020 & 0.66/0.71 & 0.40/0.63 \\
VIIRS & Tree & 0.025/0.026 & 0.97/1.01 & 0.80/0.83 \\
\midrule
Landsat 8/9 & Grass & 0.005/0.074 & 0.31/3.80 & 1.41/6.78 \\
Landsat 8/9 & Shrub & 0.020/0.034 & 0.73/1.13 & 0.37/0.97 \\
Landsat 8/9 & Tree & 0.034/0.044 & 1.27/1.56 & 1.02/1.17 \\
\midrule
Sentinel-2 & Grass & 0.011/0.055 & 0.64/2.96 & 0.64/4.53 \\
Sentinel-2 & Shrub & 0.017/0.034 & 0.58/1.16 & 0.19/0.90 \\
Sentinel-2 & Tree & 0.034/0.041 & 1.28/1.48 & 1.05/1.13 \\
\midrule
Sentinel-3 & Grass & 0.019/0.012 & 1.02/0.58 & 2.13/2.49 \\
Sentinel-3 & Shrub & 0.013/0.021 & 0.45/0.68 & 0.24/0.46 \\
Sentinel-3 & Tree & 0.015/0.017 & 0.54/0.61 & 0.11/0.63 \\
\bottomrule
\end{tabular}}
\end{table}

\vspace{-20pt}

\section{Sensitivity to Site-Suitability and LFMC-Adequacy Filters}
\label{app:filter-sensitivity}

A fixed-configuration filter sensitivity compares the final WorldCover profile
with stricter and more permissive profiles and separately removes only the
minimum LFMC standard-deviation requirement. For the stricter profile, all
products require natural vegetation fractions of at least 0.90 in the central
pixel and 0.80 across the neighbourhood, with maximum nonvegetated fraction
0.15 and a minimum fraction of 0.70 matching the central dominant class. The more permissive profile retains
the final high-resolution criteria and changes the MODIS, VIIRS and Sentinel-3
thresholds to 0.80, 0.65, 0.30 and 0.55, respectively. These comparisons retain
mean LFMC, central-pixel predictors, Random Forest settings, at least 20 field
dates and a 365-day record span. Product-specific quality screening is unchanged.
Each cohort is reconstructed from the original sampling table; acquisition-blocked
folds and event weights are generated for that cohort, with calibration and
model preprocessing estimated within each training fold. Changes in eligibility
can alter evaluation distributions and folds, so the resulting scores describe
cohort sensitivity rather than paired filter effects. The production filtering
criteria are retained. Full results and eligibility audits are in Supplementary
File S5.

Filter sensitivity results in Supplementary File S5 quantify the effect of
cohort definition under fixed central-pixel model configurations. For Grass,
the stricter, final and more permissive WorldCover profiles retain 84, 89 and
97 sites, respectively. For Shrub, the stricter, final and more permissive
WorldCover profiles retain 254, 274 and 287 sites, respectively. For Tree, the
stricter, final and more permissive WorldCover profiles retain 115, 125 and
128 sites, respectively. The absolute changes relative to the current profile
are at most 0.016 in $R^2$ and 0.49 LFMC percentage points in RMSE. Removing
only the LFMC variability cutoff expands the eligible records while retaining
the observation-count and temporal-span requirements. Grass then includes 93
sites and yields $R^2=0.712$ and RMSE 44.2 LFMC percentage points. Shrub then
includes 398 sites and yields $R^2=0.686$ and RMSE 21.2 LFMC percentage
points. Tree then includes 678 sites and yields $R^2=0.534$ and RMSE 21.2 LFMC
percentage points. These comparisons quantify sensitivity to the retained
population; they do not provide an untouched evaluation after model-family
development~\mbox{selection}.

\section{Optical-Free Sentinel-1 LFMC Retrieval}
\label{app:sentinel1-optical-free}

This complementary diagnostic evaluates LFMC retrieval without requiring an
optical satellite observation. Sentinel-1A/B interferometric wide-swath
VV/VH ground-range detected observations are matched directly to GlobeLFMC
field dates, independently of optical availability. The nearest eligible
acquisition within the prescribed $\pm$7-day calendar window is sampled on a
10~m pixel-spacing grid. Central-pixel predictors comprise VV and VH in dB,
the VH/VV ratio in dB (VH minus VV), and incidence angle. Radar quality
screening requires $-$50 < VV < 10~dB,
$-$55 < VH < 5~dB and incidence angles of \mbox{29--46 degrees}.
At least 75\% of the surrounding 3-by-3 pixels must be valid. Site suitability
requires a natural central WorldCover 2021 class, at least 80\% natural
vegetation within the surrounding footprint, at most 15\% nonvegetated cover
and at least 70\% cover matching the central class. These rules select the
field-matched modelling cohort; the neighbourhood summaries are not LFMC
predictors.

Field-quality rules and class-specific LFMC bounds follow the main analysis.
Retained measurements are deduplicated and averaged once within each site,
date and functional class. Each evaluated field event has unit weight.
Following complete-predictor screening, site/class records must have at least
20 distinct field dates spanning at least 365 days and LFMC standard deviations
of at least 20 percentage points for Grass and Shrub and 30 for Tree.
The resulting cohorts contain 1803 events at 43 Grass sites, 11,219 events at
211 Shrub sites and 4127 events at 75 Tree sites. All evaluation dates are
between 2014 and 2023. The represented countries are the USA, France, Spain
and Australia.

Two fixed Random Forest branches use identical events, targets, five
within-site outer folds and training weights. The environmental-only baseline
uses weather, topography and cyclic day of year; the radar branch adds the four
central-pixel Sentinel-1 predictors. RF settings match
Table~\ref{tab:model-hyperparameters}; central-pixel predictors and arithmetic-mean
LFMC are fixed, with no selection among footprint statistics, targets or model
families. Reused Sentinel-1 acquisitions at each site are kept in one fold,
and every evaluated site remains represented in training. Correlation pruning,
scaling, LFMC-bin training weights and target transformations are fitted within
the training folds. Paired uncertainty is estimated from 10,000 whole-site
bootstrap resamples of the out-of-fold predictions. This design evaluates
additional dates at represented sites and does not evaluate transfer to unseen
sites or regions. The cohort differs from the optical multisensor cohort, so
its scores cannot be used as a paired radar-versus-optical comparison.

Table~\ref{tab:sentinel1-optical-free} reports the paired results. Adding radar
changes $R^2$ by $-$0.016 for Grass, $-$0.005 for Shrub and +0.001 for Tree,
with RMSE changes of +0.90, +0.16 and $-$0.02 LFMC percentage points,
respectively. The paired intervals favour the environmental-only baseline
for Grass and Shrub and include zero for Tree. These results support the
feasibility of an optical-free predictor configuration but do not establish
an incremental predictive benefit from its radar predictors.

Table~\ref{tab:sentinel1-importance} summarises predictor-group diagnostics from
the outer-fold radar models. Weather accounts for the largest mean impurity
importance in all three classes; terrain has a substantial contribution for
Grass, while seasonality is more prominent for Tree. The radar group accounts
for 11.5\%, 9.9\% and 6.8\% for Grass, Shrub and Tree, respectively. Jointly
permuting the retained radar predictors across held-out observations lowers
$R^2$ by an average of 0.042, 0.018 and 0.020. Thus the fitted radar models use
these inputs, although the refitted baseline comparison shows no corresponding
accuracy gain. Permutation changes the predictor relationships presented to
an already fitted model; it addresses a different question from removing
radar and refitting the model. Correlated predictors and interactions also
prevent interpreting either importance diagnostic as an independent or causal
contribution. Supplementary File S7 provides complete paired predictions,
fold support, performance intervals, individual-feature importance and grouped
permutation~\mbox{results}.

\begin{table}[H]
\caption{Optical-free Sentinel-1 comparison with an environmental-only baseline on identical field events and acquisition-blocked folds. Both fixed Random Forest branches use weather, topography, cyclic day of year and mean LFMC; the S1 branch adds central-pixel VV, VH, VH/VV in dB and incidence angle. Differences are S1 minus environmental-only performance, with paired 95\% intervals from 10,000 whole-site bootstrap resamples. RMSE is in LFMC percentage points.}
\label{tab:sentinel1-optical-free}

\begin{adjustwidth}{-\extralength}{0cm}
\resizebox{\fulllength}{!}{
\begin{tabular}{lrcccccc}
\toprule
\textbf{Vegetation} & \textbf{Events/Sites} & \boldmath{$R^2$} \textbf{Env.} & \boldmath{$R^2$} \textbf{S1} & \boldmath{$\Delta R^2$} \textbf{[95\% CI]} & \textbf{RMSE Env.} & \textbf{RMSE S1} & \boldmath{$\Delta$}\textbf{RMSE [95\% CI]} \\
\midrule
Grass & 1803/43 & 0.587 & 0.571 & $-$0.016 [$-$0.033, $-$0.004] & 45.4 & 46.3 & 0.90 [0.19, 1.69] \\
Shrub & 11,219/211 & 0.632 & 0.628 & $-$0.005 [$-$0.009, $-$0.001] & 25.3 & 25.4 & 0.16 [0.02, 0.31] \\
Tree & 4127/75 & 0.670 & 0.670 & 0.001 [$-$0.006, 0.008] & 26.7 & 26.7 & $-$0.02 [$-$0.30, 0.26] \\
\bottomrule
\end{tabular}}
\end{adjustwidth}
\end{table}

\vspace{-20pt}

\begin{table}[H]
\caption{Predictor-group importance in the optical-free Sentinel-1 Random Forest branch. Impurity importance is summed within groups and averaged over five outer-fold fits, expressed as percentages summing to 100\% within each class. The final column reports the mean decrease in held-out $R^2$ after joint permutation of the retained radar predictors (three repeats per fold). Permutation results measure dependence of the fitted model on a group; they do not measure the gain from adding that group to a refitted baseline.}
\label{tab:sentinel1-importance}

\begin{adjustwidth}{-\extralength}{0cm}
%\centering %% If there is a figure in wide page, please release command \centering, for Table, ``\textwidth" should be ``\fulllength"
\begin{tabularx}{\fulllength}{LCccCc}
\toprule
\textbf{Vegetation} & \textbf{Weather (\%)} & \textbf{Topography (\%)} & \textbf{Seasonality (\%)} & \textbf{Radar (\%)} & \textbf{Radar Permutation} \boldmath{$\Delta R^2$} \\
\midrule
Grass & 44.4 & 30.5 & 13.6 & 11.5 & 0.042 \\
Shrub & 52.0 & 18.9 & 19.2 & 9.9 & 0.018 \\
Tree & 61.4 & 8.7 & 23.1 & 6.8 & 0.020 \\
\bottomrule
\end{tabularx}
\end{adjustwidth}
\end{table}

The optical-free configuration may be useful where persistent cloud cover
interrupts the availability and temporal continuity of optical observations.
Future work could test physics-informed Sentinel-1 models that constrain the
relationships between radar backscatter, vegetation water content and canopy
structure, building on the physics-assisted approach of Rao et~al.~\citep{rao2020sar}. Such models should be evaluated against the environmental-only
baseline on matched observations to assess whether the radar information
improves LFMC~\mbox{retrieval}.

\begin{adjustwidth}{-\extralength}{0cm}
\reftitle{References}

\PublishersNote{}
\end{adjustwidth}
\end{document}